\documentclass[a4paper, 10pt, twocolumn]{article}

\usepackage{fullpage}

\usepackage[dvipsnames]{xcolor}
\definecolor{mydarkblue}{rgb}{0,0.08,0.45}
\usepackage{hyperref}
\hypersetup{colorlinks,linkcolor={mydarkblue},citecolor={mydarkblue},urlcolor={mydarkblue}}
\usepackage{graphicx}
\usepackage[labelformat=simple]{subcaption}

\usepackage{multirow, multicol}
\usepackage{booktabs}

\usepackage{amsthm}
\usepackage{bbm}
\newtheorem{theorem}{Theorem}
\newtheorem{definition}{Definition}
\newtheorem{corollary}[theorem]{Corollary}

\newtheorem{proposition}{Proposition}

\usepackage[round]{natbib}

\usepackage{amsmath,amsfonts,bm}

\newcommand{\renyi}{R$\acute{\text{e}}$nyi}

\def\figref#1{figure~\ref{#1}}
\def\Figref#1{Figure~\ref{#1}}

\def\TwoFigref#1#2{Figures \ref{#1} and \ref{#2}}

\def\secref#1{section~\ref{#1}}

\def\eqref#1{equation~\ref{#1}}

\def\eqrefp#1{equation~(\ref{#1})} 

\def\thmref#1{theorem~\ref{#1}}
\def\Thmref#1{Theorem~\ref{#1}}
\def\Corollref#1{Corollary~\ref{#1}}  
\def\tableref#1{Table~\ref{#1}}  
\def\Propref#1{Proposition~\ref{#1}}

\newcommand{\explainEq}[1]{\overset{\underset{\mathrm{(#1)}}{}}{=}}
\newcommand{\explainLeq}[1]{\overset{\underset{\mathrm{(#1)}}{}}{\leq}}

\newcommand{\explainGeq}[1]{\overset{\underset{\mathrm{(#1)}}{}}{\geq}}

\def\1{\bm{1}}

\def\eps{{\epsilon}}

\DeclareMathAlphabet{\mathsfit}{\encodingdefault}{\sfdefault}{m}{sl}
\SetMathAlphabet{\mathsfit}{bold}{\encodingdefault}{\sfdefault}{bx}{n}

\newcommand{\E}{\mathbb{E}}

\DeclareMathOperator*{\argmax}{arg\,max}

\DeclareMathOperator{\Tr}{Tr}

\title{Generating private data with user customization}

\author{Xiao Chen, Thomas Navidi, Ram Rajagopal \\
Stanford University \\
\texttt{\{markcx, tnavidi, ramr\}@stanford.edu}}

\date{}

\begin{document}

%

%





\maketitle

\begin{abstract}
  Personal devices such as mobile phones can produce and store large amounts of data that can enhance machine learning models; however, this data may contain private information specific to the data owner that prevents the release of the data. We want to reduce the correlation between user-specific private information and the data while retaining the useful information. Rather than training a large model to achieve privatization from end to end, we first decouple the creation of a latent representation, and then privatize the data that allows user-specific privatization to occur in a setting with limited computation and minimal disturbance on the utility of the data. We leverage a Variational Autoencoder (VAE) to create a compact latent representation of the data that remains fixed for all devices and all possible private labels. We then train a small generative filter to perturb the latent representation based on user specified preferences regarding the private and utility information. The small filter is trained via a GAN-type robust optimization that can take place on a distributed device such as a phone or tablet. Under special conditions of our linear filter, we disclose the connections between our generative approach and \renyi~differential privacy. We conduct experiments on multiple datasets including MNIST, UCI-Adult, and CelebA, and give a thorough evaluation including visualizing the geometry of the latent embeddings and estimating the empirical mutual information to show the effectiveness of our approach.     
\end{abstract}

\section{Introduction}
The success of machine learning algorithms relies on not only technical methodologies, but also the availability of large datasets such as images \citep{krizhevsky2012imagenet}; however, data can often contain sensitive information, such as race or age, that may hinder the owner's ability to release the data to grasp its utility. We are interested in exploring methods of providing privatized data such that sensitive information cannot be easily inferred from the adversarial perspective, while preserving the utility of the dataset. In particular, we consider a setting where many end users are independently gathering data that will be collected by a third party. Each user is incentivized to label their own data with useful information; however, they have the option to create private labels for information that they do not want to share with the database. In the case where data contains a large amount of information such as images, there can be an overwhelming number of potential private and utility label combinations (skin color, age, gender, race, location, medical conditions, etc.). The large number of combinations prevents training a separate method to obscure each set of labels centrally. Furthermore, when participants are collecting data on their personal devices such as mobile phones, they would like to remove private information before the data leaves their devices. Both the large number of personal label combinations coupled with the use of mobile devices requires a privacy scheme to be computationally efficient. In this paper, we propose a method of generating private datasets that makes use of a fixed encoding, thus requiring only a few small neural networks to be trained for each label combination. This approach allows data collecting participants to select any combination of private and utility labels and remove them from the data on their own mobile devices before sending any information to a third party.

In the context of publishing datasets with privacy and utility guarantees, we briefly review a number of similar approaches that have been recently considered, and discuss why they are inadequate at performing in the distributed and customizable setting we have proposed. Traditionally in the domain of generating private datasets, researchers have made use of differential privacy (DP)\citep{dwork2006calibrating}, which involves injecting certain random noise into the data to prevent the identification of sensitive information \citep{dwork2006calibrating, dwork2011differential, dwork2014algorithmic}. However, finding a globally optimal perturbation using DP may be too stringent of a privacy condition in many high-dimensional data applications. In more recent literature, researchers commonly use Autoencoders \citep{kingma2013auto} to create a compact latent representation of the data, which does not contain private information, but does encode the useful information \citep{edwards2015censoring, abadi2016learning, beutel2017data, madras2018learning, song2018learning, chen2018understanding}. A few papers combine strategies involving both DP and Autoencoders \citep{hamm2017minimax, liu2017deeprotect}; however, all of these recent strategies require training a separate Autoencoder for each possible combination of private and utility labels. Training an Autoencoder for each privacy combination can be computationally prohibitive, especially when working with high dimensional data or when computation must be done on a small local device such as a mobile phone. Therefore, such methods are unable to handle the scenario where each participant must locally train a data generator that obscures their individual choice of private and utility labels. We believe reducing the computation and communication burden is important when dealing with distributed data, because this would be beneficial in many applications such as federated learning \citep{mcmahan2016communication}.

Another line of studies branching on differential privacy focuses on theoretical privacy guarantees of privatization mechanisms. We leverage a previously established relaxation of differential privacy along this line of work, i.e. \renyi~differential privacy \citep{dwork2016concentrated, bun2016concentrated, mironov2017renyi}, to determine how much privacy our mechanism can achieve under certain conditions. This notion of differential privacy is weaker than the more common relaxation of $(\varepsilon, \delta)$-differential privacy \citep{dwork2011differential}. 

Primarily, this paper introduces a decoupling of the creation of a latent representation and the privatization of data that allows the privatization to occur in a setting with limited computation and minimal disturbance on the utility of the data. We leverage a generative linear model to create a privatized representation of the data and establish the connection between this simple linear transformation of generative noise with differential privacy. We also build a connection between solving constrained, robust optimization and having \renyi~differential privacy under certain conditions. Finally, we run thorough empirical experiments on realistic high-dimensional datasets with comparison to the related studies. Additionally we contribute a variant on the Autoencoder to improve robustness of the decoder for reconstructing perturbed data and comprehensive investigations into: (i) the latent geometry of the data distribution before and after privatization, (ii) how training against a cross-entropy loss adversary impacts the mutual information between the data and the private label, and (iii) how our linear filter affects the classification accuracy of sensitive labels. 


\section{Problem Statement and Methodology}
Inspired by a few recent studies \citep{louizos2015variational, huang2017context, madras2018learning}, we consider the data privatization as a game between two players: the data generator (data owner) and the adversary (discriminator). The generator tries to inject noise into the data to privatize certain sensitive information contained in the data, while the adversary tries to infer this sensitive information from the data. In order to deal with high dimensional data, we first learn a latent representation or manifold of the data distribution, and then inject noise with specific latent features to reduce the correlation between released data and sensitive information. After the noise has been added to the latent vector, the data can be reconstructed and published without fear of releasing sensitive information. To summarize, the input to our system is the original dataset with both useful and private labels, and the output is a perturbed dataset that has reduced statistical correlation with the private labels, but has maintained information related to the useful labels.

\begin{figure}[!hpbt]
    \centerline{
    \includegraphics[width=0.49\textwidth]{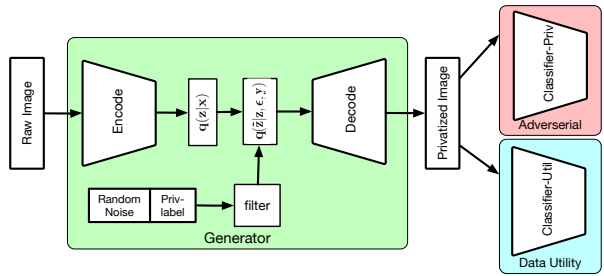}
    }
    \caption{Privatization architecture. We decompose the privatization procedure into two steps: 1) training an encoder and decoder; 2) learning a generative filter. The generative filter is learned through a min-max optimization that minimizes utility classifier loss and maximizes adversarial classifier loss. }
    \label{fig:learn:schematic:architecture}
\end{figure}

We consider the general setting where a data owner holds a dataset $\mathcal{D}$ that consists of original data $X$, private/sensitive labels $Y$, and useful labels $U$. Thus, each sample $i$ has a record $(x_i, y_i, u_i) \in \mathcal{D}$. We denote the function $g$ as a general mechanism for perturbing the data that enables owner to release the data. The released data is denoted as $\{\tilde{X}, \tilde{U}\}$. Because $Y$ is private information, it won't be released. Thus, for the record $i$, the released data can be described as $(\tilde{x}_i, \tilde{u}_i) = g(x_i, y_i, u_i).$ We simplify the problem by considering only the case that $\tilde{x}_i = g(x_i, y_i)$\footnote{We maintain the $U$ to be unchanged} for the following description: The corresponding perturbed data $\tilde{X} = g(X, Y)$ and utility attributes $U$ are published for use. The adversary builds a learning algorithm $h$ to infer the sensitive information given the released data, i.e. $ \hat{Y} = h(\tilde{X})$ where $\hat{Y}$ is the estimate of $Y$. The goal of the adversary is to minimize the inference loss $\ell{(\hat{Y}, Y)} = \ell{\Big(h\big(g(X, Y)\big), Y\Big)}$ on the private labels. Similarly, we denote the estimate of utility labels as $\hat{U} = \nu(\tilde{X}) = \nu(g(X, Y)) $. We quantify the utility of the released data through another loss function $\tilde{\ell}$ that captures the utility attributes, i.e. $\tilde{\ell}(\tilde{X}, U ) = \tilde{\ell}\Big(\nu\big(g(X, Y)\big), U \Big)$. The data owner wants to maximize the loss that the adversary experiences in order to protect the sensitive information while maintaining the data utility by minimizing the utility loss. Given the previous settings, the data-releasing game can be expressed as follows: 
\begin{align*}
    \max_{g \in \mathcal{G}}\Big\{ \min_{h \in \mathcal{H}}\mathbb{E}\Big[ \ell{\Big(h\big(g(X, Y)\big), Y\Big)}\Big]  \\ 
    - \beta \min_{\nu \in \mathcal{V}}  \mathbb{E}\Big[\tilde{\ell}\Big(\nu\big( g(X, Y) \big), U \Big)\Big]  \Big\},
\end{align*}
where $\beta$ is a hyper-parameter weighing the trade-off between different losses, and the expectation $\mathbb{E}$ is taken over all samples from the dataset. The loss functions in this game are flexible and can be tailored to a specific metric that the data owner is interested in. For example, a typical loss function for classification problems is cross-entropy loss \citep{de2005tutorial}. Because optimizing over the functions $g, h, \nu$ is hard to implement, we use a variant of the min-max game that leverages neural networks to approximate the functions. The foundation of our approach is to construct a good posterior approximation for the data distribution in latent space $\mathcal{Z}$, and then to inject context aware noise through a filter in the latent space, and finally to run the adversarial training to achieve convergence, as illustrated in
\Figref{fig:learn:schematic:architecture}. Specifically, we consider the data owner playing the generator role that comprises a Variational Autoencoder (VAE) \citep{kingma2013auto} 
structure with an additional noise injection filter in the latent space. We use $\theta_g$, $\theta_h$, and $\theta_{\nu}$ to denote the parameters of neural networks that represent the data owner (generator), adversary (discriminator), and utility learner (util-classifier), respectively. Moreover, the parameters of generator $\theta_g$ consists of the encoder parameters $\theta_e$, decoder parameters $\theta_d$, and filter parameters $\theta_f$. The encoder and decoder parameters are trained independently of the privatization process and left fixed, since we decoupled the steps of learning latent representations and generating the privatized version of the data. Hence, we have 
\begin{align}
\begin{split}
     & \max_{\theta_f}\Big\{ \min_{\theta_h}\mathbb{E}\Big[\ell{\Big(h_{\theta_h}\big(g_{\theta_g}(X, Y)\big), Y\Big)}\Big] \\
     & - \beta \min_{\theta_{\nu}} \mathbb{E}\Big[\tilde{\ell} \Big(\nu_{\theta_{\nu}}\big( g_{\theta_g}(X, Y) \big), U \Big)\Big]  \Big\} 
\end{split}     \\
     s.t. &  \qquad D\Big(g_{\theta_e}(X), g_{\theta_f}\big(g_{\theta_e}(X), \eps, Y\big)\Big) \leq b \label{eq:main:frame:budget:c1},
\end{align}
where $\eps$ is standard Gaussian noise, $D$ is a distance or divergence measure, and $b$ is the corresponding distortion budget. The purpose of adding $\eps$ is to randomize the noise injection generation. The divergence measure captures the closeness of the privatized samples to the original samples while the distortion budget provides a limit on this divergence. The purpose of the distortion budget is to prevent excessive noise injection, which is to avoid deteriorating  unspecified information in the data.

In principle, we perform the following three steps to complete each experiment.

1) \textbf{Train a VAE} for the generator without noise injection or min-max adversarial training. Because we want to learn a compact latent representation of the data. Rather than imposing the distortion budget at the beginning, we first train the following objective 
\begin{align}
\begin{split}
    & \min_{\theta_e, \theta_d}  -\mathbb{E}_{q({z}|{x};\theta_e)} [\log p({x}|{z};\theta_d)]  \\
     & + D_{\text{KL}}\big(q({z}|{x};\theta_e) || p({z})\big), \label{eq:vae:nelbo}
\end{split}
\end{align} 
where the posterior distribution $q(z|x; \theta_e)$ is characterized by an encoding network $g_{\theta_e}(x)$, and $p(x|z; \theta_d)$ is similarly the decoding network $g_{\theta_d}(z)$. The distribution $p(z)$ is a prior distribution that is usually chosen to be a multivariate Gaussian for the reparameterization purpose \citep{kingma2013auto}. When dealing with high dimensional data, we develop a variant of the preceding objective that captures three items: the reconstruction loss, KL divergence on latent representations, and improved robustness of the decoder network to perturbations in the latent space (as shown in \eqrefp{eq:supp:vae:loss:variant}). We discuss more details of training a VAE and this variant in \secref{supp:tech:approch:vae:explained}.

2) \textbf{Formulate the robust optimization} for min-max GAN-type training \citep{goodfellow2014generative} with noise injection, which comprises a linear filter,\footnote{ The filter can be nonlinear such as a small neural network. We focus on the linear case for the remainder of the paper.} while freezing the weights of the encoder and decoder. In this phase, we instantiate several divergence metrics and various distortion budgets to run our experiments (details in section \ref{supp:tech:approach:noise_inject:train}). When we fix the encoder, the latent variable $z$ can be expressed as $z=g_{\theta_e}(x)$ (or $z \sim q_{\theta_e}(z|x)$ or $q_{\theta_e}$ for short), and the new altered latent representation $\tilde{z}$ can be expressed as $\tilde{z}=g_{\theta_f}(z, \epsilon, y)$, where $g_{\theta_f}$ represents a filter function (or $\tilde{z} \sim q_{\theta_f}$ for short). The classifiers $h$ and $\nu$ can take the latent vector as input when training the filter to reduce the computational burden, as is done in our experiments. We focus on a canonical form for the adversarial training and cast our problem into the following robust optimization problem: 
\begin{align}
    & \min_{\theta_{h}}\Big\{ \max_{q_{\theta_f}}\mathbb{E}_{q_{\theta_f}}\big(\ell(h_{\theta_h}; \tilde{z}) \big) - \beta\mathbb{E}_{q_{\theta_f}}\big(\tilde{\ell}(\nu_{\theta_{\nu}}; \tilde{z}) \big) \Big\} \label{eq1:frame:obj} \\
    & \text{s.t.}  \quad D_f(q_{\theta_f}||q_{\theta_e}) \leq b \\
    & \qquad  \theta_{\nu} = \arg\min_{\theta_{\nu}}\mathbb{E}_{q_{\theta_f}}\big( \tilde{\ell}(\nu_{\theta_{\nu}}; \tilde{z}) \big),
\end{align}
where $D_f$ is the \emph{$f$-divergence} in this case. We omit the default condition of $\mathbb{E}q_{\theta_f} = 1$ for the simplicity of the expressions. 
%
By decomposing and analyzing some special structures of our linear filter, we disclose the connection between our generative approach and \renyi~differential privacy in \Thmref{app:theo:2} and \Propref{app:thm:prop:AA_T:bound} (in appendix \secref{supp:tech:approach:diff:privacy}).
%
%
%
%

3) \textbf{Learn the adaptive classifiers} for both the adversary and utility labels according to the data $[\tilde{X}, Y, U]$ \big(or $[\tilde{Z}, Y, U]$ if the classifier takes the latent vector as input\big), where the perturbed data $\tilde{X}$ (or $\tilde{Z}$) is generated based on the trained generator. We validate the performance of our approach by comparing metrics such as classification accuracy and empirical mutual information. Furthermore, we visualize the geometry of the latent representations, e.g. \figref{fig:circle:latent:geo:embedding:image}, to give intuitions behind how our framework achieves privacy.

\section{Experiments and Results}
\begin{figure}[!hbpt]
    \centering 
    \begin{subfigure}[t]{0.95\columnwidth}
    \includegraphics[width=1\textwidth]{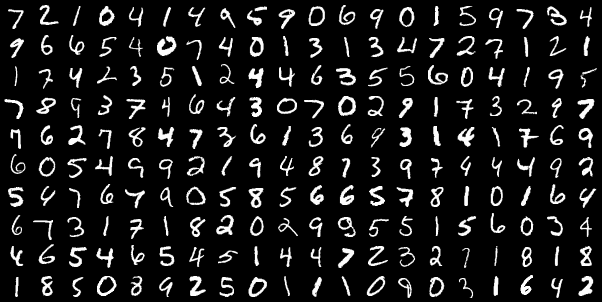}
    \caption{Sample of original images}
    \end{subfigure}
    \begin{subfigure}[t]{0.95\columnwidth}
    \includegraphics[width=1\textwidth]{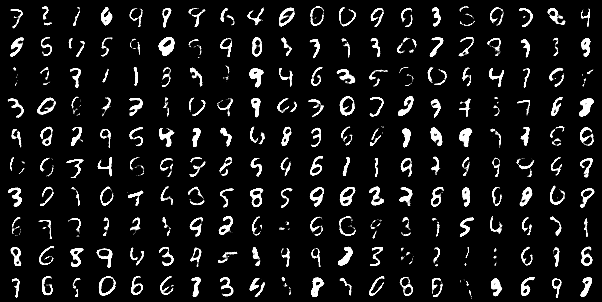}
    \caption{Same images perturbed to privatize digit ID}
    \end{subfigure}
    \caption{Visualization of digits pre- and post-noise injection and adversarial training. We find that digit IDs are randomly switched while circle digits remain circle digits and non-circle digits remain as non-circle digits. }
    \label{fig:sample_MNIST}
\end{figure}

We verified our idea on four datasets. The first is the {MNIST} dataset \citep{lecun-mnisthandwrittendigit-2010} of handwritten digits, commonly used as a toy example in machine learning literature. We have two cases involving this dataset: In MNIST Case 1, we preserve information regarding whether the digit contains a circle (i.e. digits \textsf{0,6,8,9}), but privatize the value of the digit itself. In MNIST Case 2, we preserve information on the parity (even or odd digit), but privatize whether or not the digit is greater than or equal to 5. \Figref{fig:sample_MNIST} shows a sample of the original dataset along with the same sample perturbed to remove information on the digit identity but maintain the digit as a circle-containing digit. The input to the algorithm is the original dataset with labels, while the output is the perturbed data as shown. The second experimental dataset is the UCI-adult income dataset \citep{Dua:2019} that has 45222 anonymous adults from the 1994 US Census. We preserve whether an individual has an annual income over {\$50,000} or not while privatizing the gender of that individual. The third dataset is UCI-abalone \citep{nash1994population} that consists of 4177 samples with 9 attributes including target labels. The fourth dataset is the CelebA dataset \citep{liu2015faceattributes} containing color images of celebrity faces. For this realistic example, we preserve whether the celebrity is smiling, while privatizing many different labels (gender, age, etc.) independently to demonstrate our capability to privatize a wide variety of labels with a single latent representation.

\begin{figure}[!hpbt]
    \centering
    \begin{subfigure}[t]{0.81\columnwidth}
    \includegraphics[width=1\textwidth]{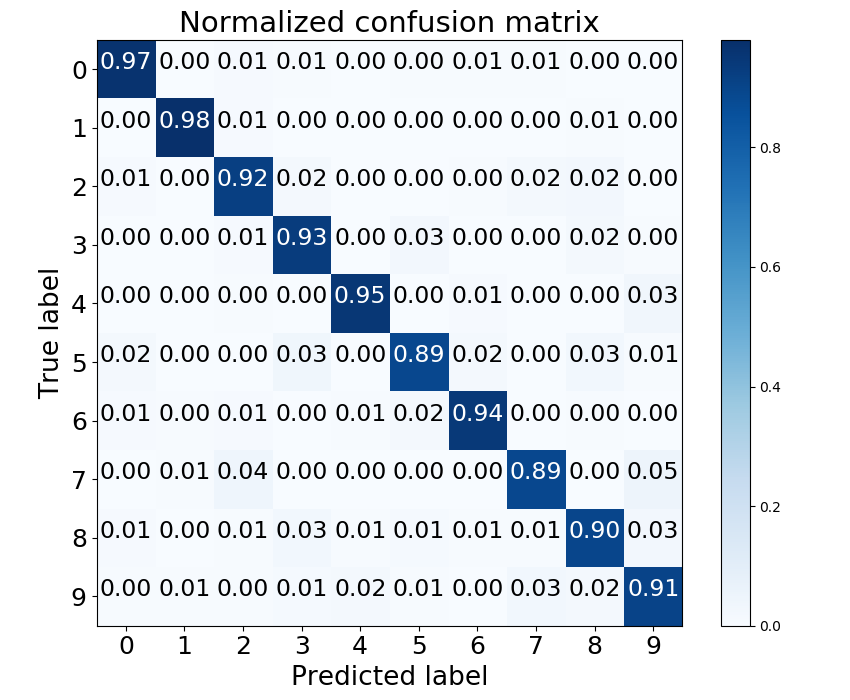}
    \caption{raw}
    \label{fig:circle:acc:confusion:mat:raw}
    \end{subfigure}
    \begin{subfigure}[t]{0.81\columnwidth}
    \includegraphics[width=1\textwidth]{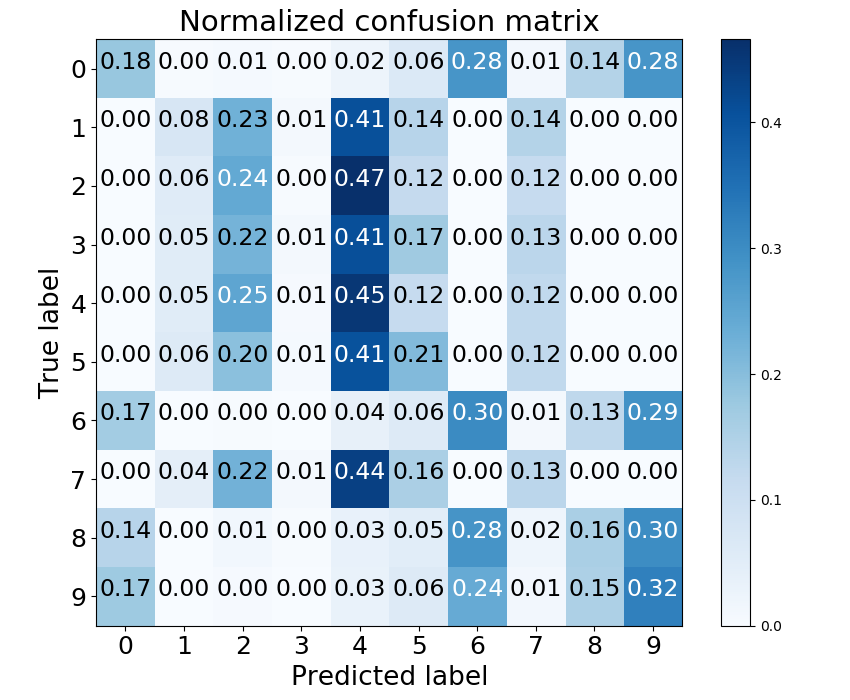}
    \caption{privatized}
    \label{fig:circle:acc:confusion:mat:perturb}
    \end{subfigure}
    \begin{subfigure}[t]{0.80\columnwidth}
    \includegraphics[width=1\textwidth]{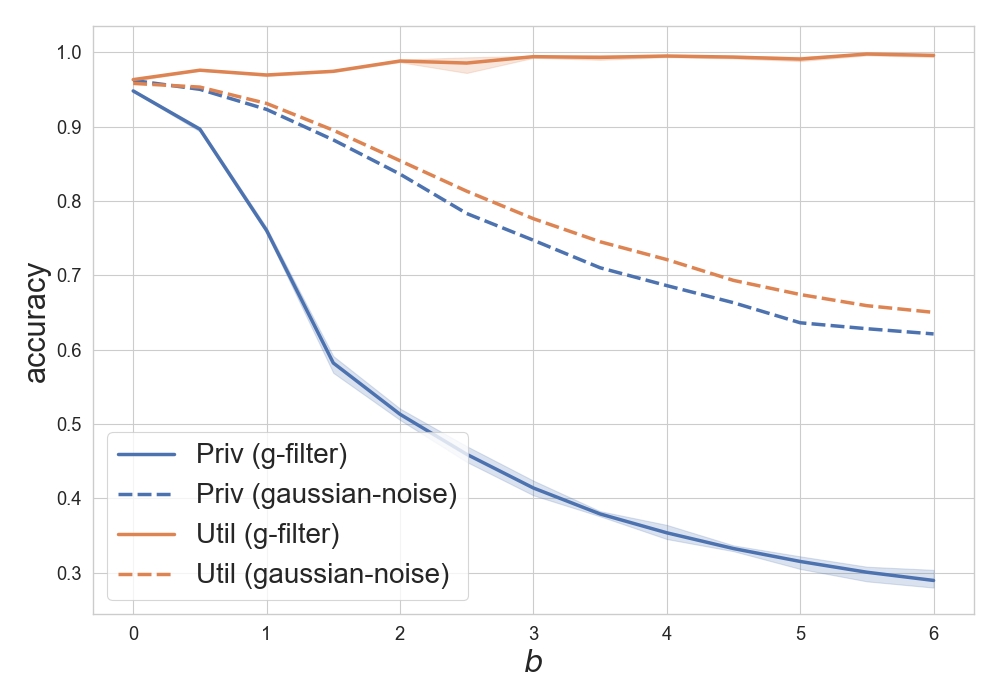}
    \caption{classification accuracy}
    \label{fig:circle:acc:distortion:compare}
    \end{subfigure}
    \caption{Classifying digits in MNIST. Original digits can be easily classified with more than 90\% accuracy on average, yet the new perturbed digits have a significantly lower accuracy as expected. Specifically, many circle digits are incorrectly classified as other circle digits and similarly for the non-circle digits. \Figref{fig:circle:acc:distortion:compare} demonstrates that classification accuracy on the private label decreases quickly while classification on the utility label remains nearly constant as the distortion budget increases. Our approach is superior to the baseline Gaussian mechanism based on the senario of adding equivalent noise on each coordinates. }
    \label{fig:mnist:circle:clf:res}
\end{figure}

\begin{figure}[!hpbt]
    \centering
    \begin{subfigure}[t]{0.81\columnwidth}
    \includegraphics[width=1\textwidth]{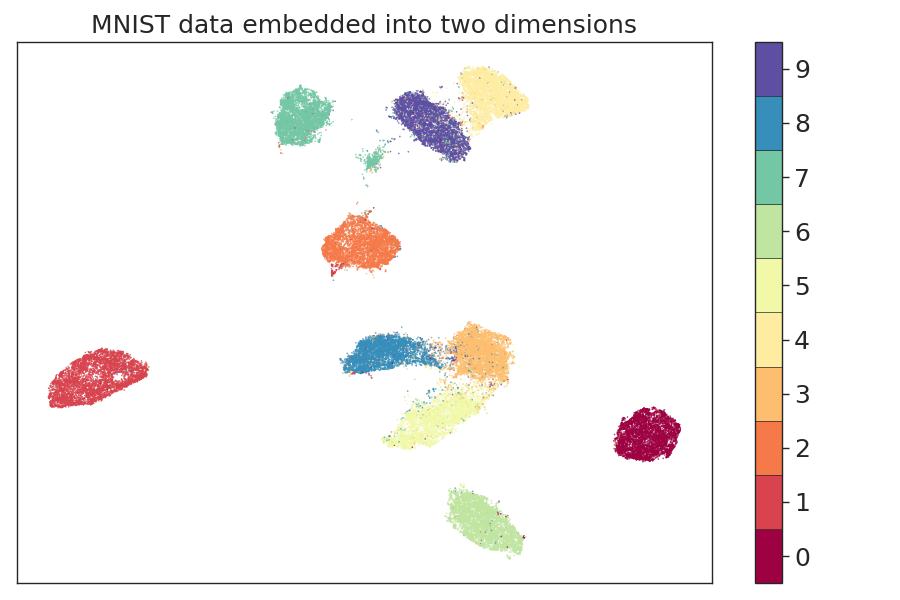}
    \caption{raw}
    \label{fig:circle:latent:geo:raw}
    \end{subfigure}
    \begin{subfigure}[t]{0.81\columnwidth}
    \includegraphics[width=1\textwidth]{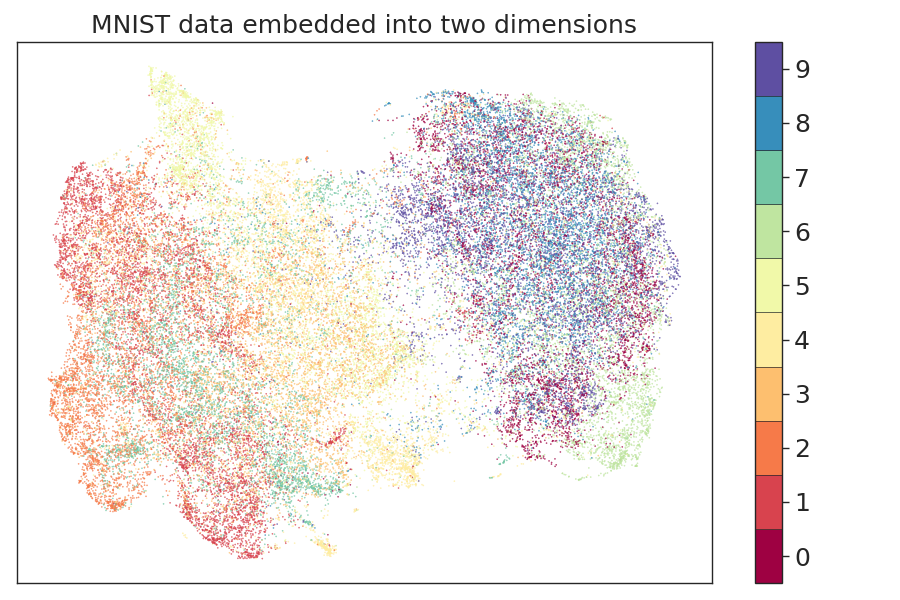}
    \caption{moderate perturbation }
    \label{fig:circle:latent:geo:moderate}
    \end{subfigure}
    \begin{subfigure}[t]{0.81\columnwidth}
    \includegraphics[width=1\textwidth]{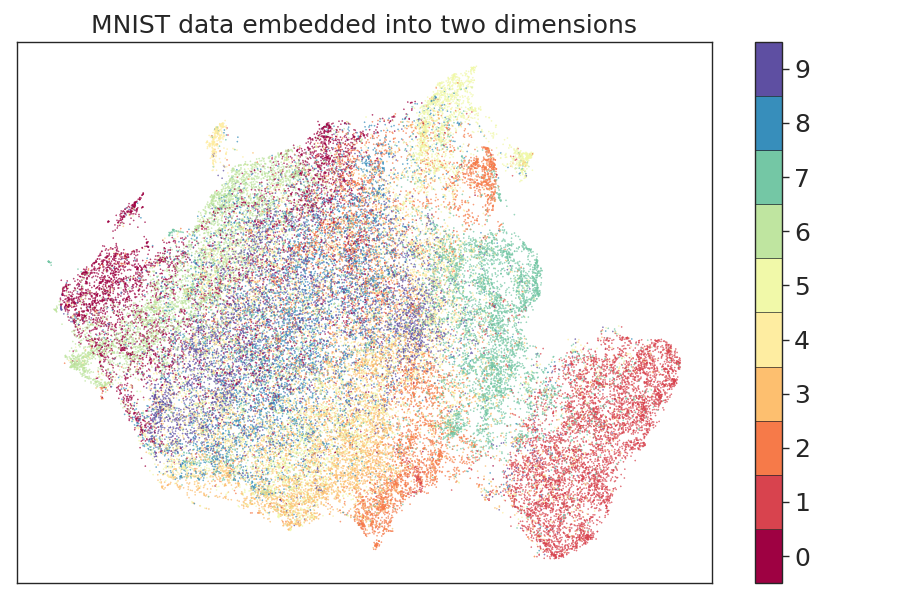}
    \caption{heavy perturbation}
    \label{fig:circle:latent:geo:heavy}
    \end{subfigure}
    \caption{Visualization of the latent geometry. The original embedding in \Figref{fig:circle:latent:geo:raw} is clearly segmented into individual clusters for each digit; however, when we allow a distortion budget of $b=1.5$, as shown in \Figref{fig:circle:latent:geo:moderate}, the digits are separated according to the circle or non-circle property by the gap between the left and right clouds of points. A larger distortion budget ($b=5$) nests all samples close together with some maintained local clusters as seen in \Figref{fig:circle:latent:geo:heavy}. }
    \label{fig:circle:latent:geo:embedding:image}
\end{figure}

\textbf{MNIST Case 1}: We considered the digit value itself as the private attribute and the digit containing a circle or not as the utility attribute. Figure~\ref{fig:sample_MNIST} shows samples of this case. Specific classification results before and after privatization are given in the form of confusion matrices in \TwoFigref{fig:circle:acc:confusion:mat:raw}{fig:circle:acc:confusion:mat:perturb}, demonstrating a significant reduction in private label classification accuracy. These results are supported by our illustrations of the latent space geometry in \Figref{fig:circle:latent:geo:embedding:image} obtained via uniform manifold approximation and projection (UMAP) \citep{mcinnes2018umap}. Specifically, \figref{fig:circle:latent:geo:moderate} shows a clear separation between circle digits (on the right) and non-circle digits (on the left). We also investigated the sensitivity of classification accuracy for both labels with respect to the distortion budget (for KL-divergence) in \Figref{fig:circle:acc:distortion:compare}, demonstrating that increasing the distortion budget rapidly decreases the private label accuracy while maintaining the accuracy of utility label. We also compare these results to a baseline method based on an additive Gaussian mechanism (discussed in \secref{supp:tech:approach:diff:privacy}), and we found that this additive Gaussian mechanism performs worse than our generative adversarial filter in terms of keeping the utility and protecting the private labels because the Gaussian mechanism yields lower utility and worse privacy (i.e. an adversary can have higher prediction accuracy of private labels) than the min-max generative filter approach. In appendix \secref{supp:tech:approach:diff:privacy}, we show our min-max generative filter approach can connect to \renyi~ differential privacy under certain conditions.         

\textbf{MNIST Case 2}: This case has the same setting as the experiment given in \cite{rezaei2018protecting} where we consider odd or even digits as the target utility and large ($\geq 5$) or small ($< 5$) value  as the private label. Rather than training a generator based on a fixed classifier, as done in \cite{rezaei2018protecting}, we take a different modeling and evaluation approach that allows the adversarial classifier to update dynamically. We find that the classification accuracy of the private attribute drops down from 95\% to 65\% as the distortion budget grows. Meanwhile our generative filter doesn't deteriorate the target utility too much, maintaining a classification accuracy above 87\% for the utility label as the distortion increases, as shown in \figref{fig:mnist:acc:inc:b:priv:geq5:util:oddeven}. We discuss more results in the appendix \secref{supp:mnist:more:res:w:mi}, together with results verifying the reduction of mutual information between the data and the private labels.

To understand whether specifying the target utility is too restrictive for other usage of the data, we conducted the following experiment. Specifically, we measure the classification accuracy of the circle attribute from case 1,  while using the privatization scheme from case 2 (preserving digit parity). This tests how the distortion budget prevents excessive loss of information of non-target attributes. When training for case 2, the circle attribute from case 1 is not included in the loss function by design; however, as seen in \tableref{tab:mnist:non-target}, the classification accuracy on the circle is not more diminished than the target attribute (odd). A more detailed plot of the classification accuracy can be found in \Figref{fig:mnist:acc:inc:b:priv:geq5:util:non-target} in the appendix \secref{supp:mnist:more:res:w:mi}. This experiment demonstrates that the privatized data maintains utility beyond the predefined utility labels used in training.

\begin{table}[!hbpt]
    \centering
    \caption{Accuracy of private label ($\geq 5$), target label (odd), and non-target label (circle) for MNIST dataset. The raw embedding yielded by the VAE before privacy is denoted as \emph{emb-raw}. The embedding yielded from the generative filter after privacy is denoted as \emph{emb-g-filter}.}
    \label{tab:mnist:non-target}
    \begin{scriptsize}
    \begin{tabular}{c|c|c|c}
    \toprule 
    \textbf{Data} & \textbf{Priv. attr.} & \textbf{Util. attr.}  & \textbf{Non-tar. attr.} \\
    \midrule 
        emb-raw & 0.951 & 0.952 & 0.951 \\
        emb-g-filter & 0.687 & 0.899 & 0.9 \\
    \bottomrule    
    \end{tabular}
    \end{scriptsize}
\end{table}

\textbf{UCI-Adult}: We conduct the experiment on this dataset by setting the private label to be gender and the utility label to be income. All the data is preprocessed to binary values for the ease of training. We compare our method with the models of Variational Fair AutoEncoder (VFAE)\citep{louizos2015variational} and Lagrangian Mutual Information-based Fair Representations (LMIFR) \citep{song2018learning}. The corresponding accuracy and area-under receiver operating characteristic curve (AUROC) of classifying private label and utility label are shown in \tableref{tab:uciadult:benchmark:res}. Our method has the lowest accuracy and the smallest AUROC on the privatized gender attribute. Although our method doesn't perform best on classifying the utility label, it still achieves comparable results in terms of both accuracy and AUROC, which are described in \tableref{tab:uciadult:benchmark:res}.

\begin{table*}[!hbpt]
    \centering
    \caption{Accuracy (acc.) and Area-Under-ROC (auroc.) of private label (gender) and target label (income) for UCI-Adult dataset.}
    \label{tab:uciadult:benchmark:res}
    \begin{small}
    \begin{tabular}{c|c|c|c|c}
    \toprule 
    \multirow{1}{*}{\textbf{Model}} & \multicolumn{2}{c|}{\textbf{Private attr. }}  & \multicolumn{2}{c}{\textbf{Utility attr. }}  \\
    & acc. & auroc. & acc. & auroc. \\ 
    \midrule 
        VAE \citep{kingma2013auto}  & 0.850 $\pm$ 0.007 & 0.843 $\pm$ 0.007 & 0.837 $\pm$ 0.009 & 0.755 $\pm$ 0.006 \\
        VFAE \citep{louizos2015variational} & 0.802 $\pm$ 0.009 & 0.703 $\pm$ 0.013 & \textbf{0.851 $\pm$ 0.007} & 0.761 $\pm$ 0.011 \\
        LMIFR \citep{song2018learning} & 0.728 $\pm$ 0.014 & 0.659 $\pm$ 0.012 & 0.829 $\pm$ 0.009 & 0.741 $\pm$ 0.013  \\
        Ours (w. generative filter) & \textbf{ 0.717 $\pm$ 0.008} & 0.632 $\pm$ 0.011 & 0.822 $\pm$ 0.011 & 0.731 $\pm$ 0.015 \\
    \bottomrule    
    \end{tabular}
    \end{small}
\end{table*}
\textbf{UCI-Abalone}: We partition the rings label into two classes based on if individual had less or more than 10 rings. Such a setup follows the same setting in \cite{jalko2016differentially}. We treat the rings label as the utility label. For the private label, we pick sex because we hope classifying rings could be less correlated with the sex of abalones. There are three categories in sex: male, female and infant. We see that having small amount of distortion budget (b=0.01) reduces the classification accuracy of private label significantly. However, the accuracy and auroc remain around the similar level comparing with the raw data, unless we have large distortion budget (b=10), shown in \tableref{tab:abalone:results}.  

\begin{figure*}[ht] 
    \centerline{
    \begin{subfigure}[t]{0.95\columnwidth}
    \includegraphics[width=1\textwidth]{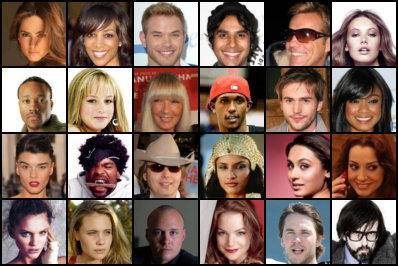}
    \caption{raw samples}
    \label{fig:celebA:samples:raw}
    \end{subfigure}
    \begin{subfigure}[t]{0.95\columnwidth}
    \includegraphics[width=1\textwidth]{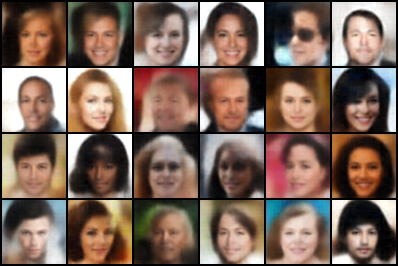}
    \caption{same samples gender privatized}
    \label{fig:celebA:samples:priv:gender}
    \end{subfigure}
    }
    \caption{Sampled images. We privatize gender while keeping the expression (smiling or not) as the utility label. The generative filter switches some female faces to male faces and also some male faces to female faces while preserving the celebrity's smile. The blurriness of the privatized images is because of the compactness of the latent representation that is generated from the VAE model, which is a normal occurrence among VAE models, and not from our privatization scheme. More details can be found in \figref{fig:celebA:raw:enc:dec:samples} in \secref{appx:sec:celebA:examples:more}.}
    \label{fig:celebA:gender:priv:examples}
\end{figure*}

\begin{table}[!bpht]
    \centering
    \caption{Accuracy (acc.) of both the private label (sex) and utility label (rings), and the Area-Under-ROC (auroc.) of utility label for the Abalone dataset. }
    \label{tab:abalone:results}
    \begin{scriptsize}
    \begin{tabular}{c|c|c|c}
    \toprule
        $b$ & \textbf{priv. attr.} & \textbf{util. attr.} & \textbf{util. attr. } \\
        & acc. & acc. & auroc. \\
    \midrule
        0 (raw) & 0.546 $\pm$ 0.011 & 0.748 $\pm$ 0.016 & 0.75 $\pm$ 0.003 \\ 
        0.01 & 0.321 $\pm$ 0.007 & 0.733 $\pm$0.010  & 0.729 $\pm$0.003 \\
        0.1 & 0.314 $\pm$ 0.010  & 0.721 $\pm$ 0.012 & 0.728 $\pm$0.003 \\
        1 & 0.313 $\pm$ 0.02 & 0.720 $\pm$  0.010 & 0.729 $\pm$0.006 \\
        10 & 0.305 $\pm$ 0.033 & 0.707 $\pm$ 0.011 & 0.699 $\pm$0.009 \\
    \bottomrule
    \end{tabular}
    \end{scriptsize}
\end{table}
%
%
\textbf{CelebA}: For the CelebA dataset, we consider the case when there exist many private and utility label combinations depending on the user's preferences. Specifically, we experiment on the private labels gender (male), age (young), attractive, eyeglasses, big nose, big lips, high cheekbones, or wavy hair, and we set the utility label as smiling for each private label to simplify the experiment. \tableref{tab:celebA:clf:acc:results} shows the classification results of multiple approaches. Our trained generative adversarial filter reduces the accuracy down to 73\% on average, which is only 6\% more than the worst case accuracy demonstrating the ability to protect the private attributes. Meanwhile, we only sacrifice a small amount of the utility accuracy (3\% drop), which ensures that the privatized data can still serve the desired classification tasks. (All details are summarized in \tableref{tab:celebA:clf:acc:results}.) We show samples of the gender-privatized images in \Figref{fig:celebA:gender:priv:examples}, which indicates the desired phenomenon that some female images are switched into male images and some male images are changed into female images. More example images on other privatized attributes, including eyeglasses, wavy hair, and attractive, can be found in appendix \secref{appx:sec:celebA:examples:more}.
%
%
\begin{table*}[!bpht]
\caption{Classification accuracy on CelebA. The row \textbf{VAE-emb} is our initial classification accuracy on the latent vectors from our trained encoder. The row \textbf{Random-guess} demonstrates the worst possible classification accuracy. The row \textbf{VAE-g-filter} is the classification accuracy on the perturbed latent vectors yielded by our generative adversarial filters. The state-of-the-art classifier \citep{torfason2016face} can achieve 87\% accuracy on the listed private labels on average while our trained VAE can achieve a comparable accuracy (83\% on average) on the more compact latent representations. More importantly, our trained generative adversarial filter can reduce the accuracy down to 73\% on average, which is only 6\% more than the worst case accuracy demonstrating the ability to protect the private attributes. Meanwhile, we only sacrifice a small amount of the utility accuracy (3\% drop), which ensures that the privatized data can still serve the desired classification tasks}
\label{tab:celebA:clf:acc:results}
    \centerline{
    \begin{scriptsize}
    \begin{tabular}{c|cccccccc|c|c}
    \toprule
    & \multicolumn{9}{c|}{\textbf{Private attr.}} & \textbf{Utility attr.}\\
         & Male & Young & Attractive & H. Cheekbones & B. lips & B. nose & Eyeglasses & W. Hair & \textbf{Avg} & Smiling \\
    \midrule      
        \cite{liu2015faceattributes} & 0.98 & 0.87 & 0.81 & 0.87 & 0.68 & 0.78 & 0.99 & 0.80 &   0.84  &  0.92 \\ 
        \cite{torfason2016face} & 0.98 & 0.89 & 0.82 & 0.87 & 0.73 & 0.83 & 0.99 & 0.84 &  0.87 & 0.93 \\
        VAE-emb & 0.90 & 0.84 & 0.80 & 0.85 & 0.68 & 0.78 & 0.98 & 0.78 &    0.83 & 0.86 \\
        Random-guess & 0.51  &  0.76 & 0.54  &  0.54 & 0.67 & 0.77 & 0.93  &0.63 &   0.67 & 0.50 \\
        VAE-g-filter & 0.61 & 0.76 & 0.62 & 0.78 & 0.67 & 0.78 & 0.93 & 0.66 & 0.73 &   0.83 \\
        \bottomrule 
    \end{tabular}
    \end{scriptsize}
    }
\end{table*}

\section{Discussion}
\textbf{Distributed training with customized preferences}: In order to clarify how our scheme can be run in a local and distributed fashion, we performed a basic experiment on the MNIST dataset with 2 independent users to demonstrate this capability. The first user adopts the label of digit $\geq 5$ as private and odd or even as the utility label. The second user prefers the opposite and wishes to privatize odd or even and maintain $\geq 5$ as the utility label. Details on the process can be found in appendix \secref{supp:distributed_details}. 


The VAE is trained separately by a data aggregator and the parameters are handed to each user. Then, each user learns a separate linear generative filter that privatizes their data. Since the linear generative filter is trained on the low dimensional latent representation, it is small enough for computation on a personal device. After privatization, we can evaluate the classification accuracy on the private and utility labels as measured by adversaries trained on the full aggregated privatized dataset, which is the combination of each users privatized data. \tableref{tab:mnist:distr} demonstrates the classification accuracy on the two users privatized data. This shows how multiple generative filters can be trained independently using a single encoding to successfully privatize small subsets of data.

\begin{table}[!pbht]
    \centering
    \caption{Accuracy of adversarial classifiers on two users private labels}
    \label{tab:mnist:distr}
    \begin{small}
    \begin{tabular}{c|c|c}
    \toprule 
    \textbf{Classifier} & \textbf{User 1} & \textbf{User 2} \\
    \textbf{type} & \textbf{(privatize $\mathbf{\geq 5}$)} & \textbf{(privatize odd)} \\
    \midrule 
        Private attr. & 0.679 & 0.651 \\
        Utility attr. & 0.896 & 0.855 \\
    \bottomrule    
    \end{tabular}
    \end{small}
\end{table}

\textbf{Connection to \renyi~differential privacy}:
We bridge the connection between our linear privatization generator and traditional differential privacy mechanisms through the following high level descriptions (proofs and details are thoroughly explained in appendix \secref{supp:tech:approach:diff:privacy}). To begin, we introduce the \renyi-divergence and a relaxation of differential privacy based on this divergence called \renyi~differential privacy \citep{mironov2017renyi} [see Definition \ref{supp:renyi_priv}]. We then in Theorem \ref{supp:thm:filter_priv} provide the specifications under which our linear filter provides $(\varepsilon, \alpha)$-\renyi~differential privacy. Finally, in Proposition \ref{app:thm:prop:AA_T:bound} we harness our robust optimization to the \renyi~differential privacy. 

In appendix \secref{supp:class_acc}, we also establish reasons why our privatization filter is able to decrease the classification accuracy of the private attributes. 

\section{Conclusion and Future Work}
In this paper, we proposed an architecture for privatizing data while maintaining the utility that decouples for use in a distributed manner. Rather than training a very deep neural network or imposing a particular discriminator to judge real or fake images, we first trained VAE that can create a comprehensive low dimensional representation from the raw data. We then found smart perturbations in the low dimensional space according to customized requirements (e.g. various choices of the private label and utility label), using a robust optimization approach. Such an architecture and procedure enables small devices such as mobile phones or home assistants (e.g. Google home mini) to run a light weight learning algorithm to privatize data under various settings on the fly. We demonstrated that our proposed additive noise method can be \renyi~differentially private under certain conditions and compared our results to a traditional additive Gaussian mechanism for differential privacy.

Finally, we discover an interesting connection between our idea of decorrelating the released data with sensitive attributes and the notion of learning fair representations\citep{pmlr-v28-zemel13, louizos2015variational,madras2018learning}. In fairness learning, the notion of demographic parity requires the prediction of a target output to be independent with respect to certain protected attributes. We find our generative filter could be used to produce a fair representation after privatizing the raw embedding, which shares a similar idea to that of demographic parity. Proving this notion along with other notions in fairness learning such as equal odds and equal opportunity \citep{hardt2016equality} will be left for future work.

\bibliography{main_paper}
\onecolumn
%
\section{Appendix: Supplementary Materials}
\subsection{Related work}
We introduce several papers \citep{hamm2017minimax, louizos2015variational, huang2017context, chen2018vgan, creager2019flexibly} which are closely related to our ideas with several distinctions. \cite{hamm2017minimax} presents a minimax filter without a decoder or a distortion constraint. It's not a generative model. The presented simulation focuses on low dimensional and time series data instead of images, so our specific model architecture, loss functions, and training details are quite different. \citet{louizos2015variational} proposed a Variational Fair Autoencoder that requires training from end to end, which is computationally expensive with high dimensional data, many privacy options, and training on edge devices. They also use Maximum Mean Discrepancy (MMD) to restrict the distance between two samples. We decouple the approach into a VAE and a linear filter, while adopting the $f$-divergence (equivalent to $\alpha$-divergence in our context) to constrain the distance between latent representations. One benefit of doing that is such a divergence has connections to \renyi~differential privacy under certain conditions. \cite{huang2017context} focuses on one-dimensional variables and uses a synthetic dataset for the simulation, which remains unclear how it can be scaled up to a realistic dataset. Many studies do not recover the privatized data to the original dimension from the latent representation to give qualitative visual support, whereas our experiments conduct a thorough evaluation through checking the latent representation and reconstructing images back to the original dimension. \cite{chen2018vgan} uses a variant of the classical GAN objective. They require the generator to take a target alternative label to privatize the original sensitive label, which leads to deterministic generation. Instead of lumping all losses together and training deep models from end to end, we decouple the encoding/decoding and noise injection without requiring a target alternative label. \cite{creager2019flexibly} proposes a framework that creates fair representation via disentangling certain labels. Although the work builds on the VAE with modifications to factorize the attributes, it focuses on the sub-group fair classification (e.g. similar false positive rate among sub-groups) rather than creating privacy-preserving data. Furthermore, we have two discriminators: private and utility. In addition to the VAE, we use KL-divergence to control the sample distortion and minimax robust optimization to learn a simple linear model. Through this, we disclose the connection to the \renyi~differential privacy, which is also a new attempt. 

\subsection{VAE training}
\label{supp:tech:approch:vae:explained}
A variational autoencoder is a generative model defining a joint probability distribution between a latent variable ${z}$ and original input ${x}$. We assume the data is generated from a parametric distribution $p({x}|{z} ; \theta)$ that depends on latent variable ${z}$, where $\theta$ are the parameters of a neural network that is usually a decoder net. Maximizing the marginal likelihood $p({x}|{z}; \theta)$ directly is usually intractable. Thus, we use the variational inference method proposed by \cite{kingma2013auto} to optimize $\log p({x}|{z}; \theta)$ over an alternative distribution $q({z}|{x}; \phi)$ with an additional KL divergence term $D_{\text{KL}}(q({z}|{x}; \phi) || p({z})\big)$ , where the $\phi$ are parameters of a neural net and $p({z})$ is an assumed prior over the latent space. The resulting cost function is often called evidence lower bound (ELBO)
\begin{align}
    \log p({x}; \theta) & \geq  \mathbb{E}_{q({z}|{x};\phi)} [\log p({x}|{z};\theta)] - D_{\text{KL}}\big(q({z}|{x};\phi) || p({z})\big) = \mathcal{L}_{\text{ELBO}} .
\end{align}
Maximizing the ELBO is implicitly maximizing the log-likelihood of $p(\mathbf{x})$. The negative objective (also known as negative ELBO) can be interpreted as minimizing the reconstruction loss of a probabilistic autoencoder and regularizing the posterior distribution towards a prior.
Although the loss of the VAE is mentioned in many studies \citep{kingma2013auto, louizos2015variational}, we include the derivation of the following relationships for context:
\begin{align}
\begin{split}
    & D_{\text{KL}}(q(z)||p(z|x; \theta)) \\
    & = \sum_{z} q(z) \log \frac{q(z)}{p(z|x;\theta)} = -\sum_{z}
q(z) \log p(z, x;\theta) + \log p(x; \theta) \underbrace{-H(q)}_{\sum_{z} q(z) \log q(z)} \geq 0
\end{split}
\end{align}.

The evidence lower bound (ELBO) for any distribution $q$ has the following property: 
\begin{align}
    & \log p(x; \theta) \geq \sum_{z}
q(z) \log p(z, x;\theta) -  \sum_{z} q(z) \log q(z)  \quad [\text{since } D_{\text{KL}}(\cdot, \cdot) \geq 0] \\
& = \sum_{z}
q(z)\Big( \log p(z, x;\theta) - \log p(z) \Big)  - \sum_{z} q(z)\Big(   \log q(z) - \log p(z)\Big) \\
& \explainEq{i} \mathbb{E}_{q(z|x; \phi)}[\log p(x|z; \theta)] - D_{\text{KL}}\big(q(z|x; \phi)||p(z)\big) = \mathcal{L}_\text{ELBO}, 
\end{align}
where equality (i) holds because we treat encoder net $q(z|x; \phi)$ as the distribution $q(z)$. By placing the corresponding parameters of the encoder and decoder networks and the negative sign on the ELBO expression, we get the loss function \eqrefp{eq:vae:nelbo}. The architecture of the encoder and decoder for the {MNIST} experiments is explained in \secref{supp:nn:archi:explain}. 
%

In our experiments with the {MNIST} dataset, the negative ELBO objective works well because each pixel value (0 black or 1 white) is generated from a Bernoulli distribution. However, in the experiments of {CelebA}, we change the reconstruction loss into the $\ell_p$ norm of the difference between the raw and reconstructed samples because the RGB pixels are not Bernoulli random variables, but real-valued random variables between 0 and 1. We still add the regularization KL term as follows:  
\begin{align}
    \mathbb{E}_{x^{\prime} \sim p({x}|{z}; \theta)}\big[||x - x^{\prime}||_{\text{p}}\big] + \gamma D_{\text{KL}}\big(q({z}|{x};\phi) || p({z})\big). 
\end{align}
Throughout the experiments we use a Gaussian $\mathcal{N}(0, {I})$ as the prior $p({z})$, $x$ is sampled from the data, and $\gamma$ is a hyper-parameter. The reconstruction loss uses the $\ell_2$ norm by default because it is widely adopted in image reconstruction, although the $\ell_1$ norm is acceptable too.

When training the VAE, we additionally ensure that small perturbations in the latent space will not yield huge deviations in the reconstructed space. More specifically, we denote the encoder and decoder to be $g_e$ and $g_d$ respectively. The generator $g$ can be considered as a composition of an encoding and decoding process, i.e. $g(X)$ = $(g_d \circ g_e) (X) = g_d(g_e(X))$, where we ignore the $Y$ inputs here for the purpose of simplifying the explanation. One desired intuitive property for the decoder is to maintain that small changes in the input latent space $\mathcal{Z}$ still produce plausible faces similar to the original latent space when reconstructed. Thus, we would like to impose some Lipschitz continuity property on the decoder, i.e. for two points $z^{(1)}, z^{(2)} \in \mathcal{Z}$, we assume $|| g_d(z^{(1)}) - g_d(z^{(2)}) || \leq C_L ||z^{(1)} - z^{(2)} ||$ where $C_L$ is some Lipschitz constant (or equivalently $||\nabla_z g_d(z)|| \leq C_L$). In the implementation of our experiments, the gradient for each batch (with size $m$) is 
\begin{align}
    \nabla_{z} g_d(z) = \begin{bmatrix} 
        \frac{\partial g_d(z^{(1)})}{\partial z^{(1)}} \\
        \vdots \\
         \frac{\partial g_d(z^{(m)})}{\partial z^{(m)}}
    \end{bmatrix} . 
\end{align}
It is worth noticing that $ 
   \frac{\partial g_d(z^{(i)})}{\partial z^{(i)}} =  \frac{\partial \sum_{i=1}^{m}g_d(z^{(i)})}{\partial z^{(i)}}$,   
because  $\frac{\partial g_d(z^{(j)})}{\partial z^{(i)}}$ = 0 when $i\neq j$. To avoid the iterative calculation of each gradient within a batch, we define $Z$ as the batched latent input, 
$
    Z = \begin{bmatrix}
    z^{(1)} \\
    \vdots \\
    z^{(m)}
    \end{bmatrix}$
and use $
    \nabla_{z} g_d(z) = \frac{\partial}{\partial Z} \sum_{i=1}^{m} g_d(z^{(i)}) 
$. The loss used for training the VAE is modified to be 
\begin{align}
    \mathbb{E}_{x^{\prime} \sim p({x}|{z}; \theta_d)}\big[||x - x^{\prime}||_{2}\big] + \gamma D_{\text{KL}}\big(q({z}|{x};\theta_e) || p({z})\big) + \kappa  \mathbb{E}_{z \sim r_{\alpha}(z)}\big[ \big(||\nabla_z (g_{\theta_d}(z)||_2 - C_L\big)_{+}^2 \big],  \label{eq:supp:vae:loss:variant}
\end{align}
\noindent where $x$ are samples drawn from the image data, $\gamma, \kappa$, and $C_L$ are hyper-parameters, and $(x)_{+}$ means $\max\{x, 0\}$. Finally, $r_\alpha$ is defined by sampling $\alpha \sim \textsf{Unif}[0,1]$,  $Z_{1} \sim p(z)$, and  $Z_{2} \sim q(z|x)$, and returning $\alpha Z_{1} + (1-\alpha) Z_{2}$. We optimize over $r_\alpha$ to ensure that points in between the prior and learned latent distribution maintain a similar reconstruction to points within the learned distribution. This gives us the Lipschitz continuity properties we desire for points perturbed outside of the original learned latent distribution.

\subsection{Robust optimization and adversarial training} \label{supp:tech:approach:noise_inject:train}
In this section, we formulate the generator training as a robust optimization. Essentially, the generator is trying to construct a new latent distribution that reduces the correlation between data samples and sensitive labels while maintaining the correlation with utility labels by leveraging the appropriate generative filters. The new latent distribution, however, cannot deviate from the original distribution too much (bounded by $b$ in \eqrefp{eq:main:frame:budget:c1}) to maintain the general quality of the reconstructed images. To simplify the notation, we will use $h$ for the classifier $h_{\theta_h}$ (a similar notion applies to $\nu$ or $\nu_{\theta_{\nu}}$). We also consider the triplet $(\tilde{z}, y, u)$ as the input data, where $\tilde{z}$ is the perturbed version of the original embedding $z$, which is the latent representation of image $x$. The values $y$ and $u$ are the sensitive label and utility label respectively. Without loss of generality, we succinctly express the loss $\ell(h; (\tilde{z}, y))$ as $\ell(h; \tilde{z})$ [similarly expressing $\tilde{\ell}(\nu; (\tilde{z}, u))$ as $\tilde{\ell}(\nu; \tilde{z})$]. We assume the sample input $\tilde{z}$ follows the distribution $q_{\psi}$ that needs to be determined. Thus, the robust optimization is   
\begin{align}
    & \min_h \max_{q_{\psi}} \mathbb{E}_{q_{\psi}}\big[ \ell(h; \tilde{z}) \big] + \beta \min_{q_{\psi}} \min_{\nu}\mathbb{E}_{q_{\psi}}\big[\tilde{\ell}(\nu; \tilde{z}) \big] \label{eq6:appx:obj} \\
    \text{s.t.} & \qquad D_f(q_{\psi}||q_{\phi} ) \leq b, 
\end{align}
where $q_{\phi}$ is the distribution of the raw embedding $z$ (also known as $q_{\theta_e}(z|x)$). In the constraint, the $f$-divergence \citep{nguyen2010estimating, cichocki2010families} between $q_{\psi}$ and $q_{\phi}$ is $D_f(q_{\psi}||q_{\phi}) = \int q_\phi(z)f(\frac{q_{\psi}(z)}{q_{\phi}(z)}) d\mu(z) $ (assuming $q_{\psi}$ and $q_{\phi}$ are absolutely continuous with respect to measure $\mu$). A few typical divergences \citep{nowozin2016f}, depending on the choices of $f$, are
\begin{enumerate}
    \item KL-divergence $D_{\text{KL}}(q_{\psi}||q_{\phi}) \leq b$, by taking $f(t)=t\log t$
    \item reverse KL-divergence $D_{\text{KL}}(q_{\phi}||q_{\psi})\leq  b$, by taking $f(t) =-\log t$
    \item $\chi^2$-divergence $D_{\chi^2}(q_{\psi}||q_{\phi}) \leq  b$, by taking $ f(t)=\frac{1}{2}(t-1)^2$. 
\end{enumerate}
In the remainder of this section, we focus on the KL and $\chi^2$ divergence to build a connection between the divergence based constraint we use and norm-ball based constraints seen in \cite{wong2018provable,madry2017towards, koh2018stronger}, and \cite{papernot2017practical}.

When we run the constrained optimization, for instance in terms of KL-divergence, we make use of the Lagrangian relaxation technique \citep{boyd2004convex} to put the distortion budget as the penalty term in the loss of objective as 
\begin{align}
    \min\big\{\lambda_1 \max\{D_{\text{KL}} - b, 0 \} + \lambda_2( D_{\text{KL}} - b)^2 \big\}. \label{eq6:ctr:relax}
\end{align}
Although this term is not necessarily convex in model parameters, the point-wise max and the quadratic term are both convex in $D_{\text{KL}}$. Such a technique is often used in many constrained optimization problems in the context of deep learning or GAN related work~\citep{gulrajani2017improved}.    

\subsubsection{Extension to multivariate Gaussian}
When we train the VAE in the beginning, we impose the distribution $q_{\phi}$ to be a multivariate Gaussian by penalizing the KL-divergence between $q_{\theta_e}(z|x)$ and a prior normal distribution $\mathcal{N}(0, {I})$, where ${I}$ is the identity matrix. Without loss of generality, we can assume the raw encoded variable $z$ follows a distribution $q_{\phi}$ that is the Gaussian distribution $\mathcal{N}(\mu_1, \Sigma_1)$ (more precisely $\mathcal{N}\big(\mu_1(x), \Sigma_1(x)\big)$, where the mean and variance depends on samples $x$, but we suppress the $x$ to simplify notation). The new perturbed distribution $q_{\psi}$ is then also a Gaussian $\mathcal{N}(\mu_2, \Sigma_2)$. Thus, the constraint for the KL divergence becomes 
\begin{align*}
    & D_{\text{KL}}(q_{\psi}||q_{\phi}) = \mathbb{E}_{q_{\psi}}(\log \frac{q_{\psi}}{q_{\phi}}) \\
    & = \frac{1}{2}\mathbb{E}_{q_{\psi}}[-\log \det(\Sigma_2) - (z - \mu_2)^T\Sigma_2^{-1}(z-\mu_2) + \log \det(\Sigma_1) + (z-\mu_1)^T\Sigma_1^{-1}(z-\mu_1) ] \\
    & = \frac{1}{2}\big[ \log\frac{\det (\Sigma_1)}{\det (\Sigma_2)} - \underbrace{d}_{=\Tr(\mathbf{I})} + \Tr(\Sigma_1^{-1}\Sigma_2) + (\mu_1 - \mu_2 )^T\Sigma_1^{-1}(\mu_1 - \mu_2) \big] \leq b.
\end{align*}
To further simplify the scenario,  we consider the case that $\Sigma_2 = \Sigma_1 = \Sigma$, then
\begin{align}
    D_{\text{KL}}(q_{\psi}||q_{\phi}) = \frac{1}{2}(\mu_1 - \mu_2)^T \Sigma^{-1} (\mu_1 - \mu_2) = \frac{1}{2} ||\mu_1 - \mu_2||_{\Sigma^{-1}}^2 \leq b \label{supp:eq:D_KL:two:gaussian}.
\end{align}
When $\Sigma={I}$, the preceding constraint is equivalent to $\frac{1}{2}||\mu_1 - \mu_2 ||_2^2$. It is worth mentioning that such a divergence-based constraint is also connected to the norm-ball based constraint on samples.

In the case of $\chi^2$-divergence,
\begin{align*}
    & D_{\chi^2}(q_{\psi}||q_{\phi})  = E_{q_{\phi}}[\frac{1}{2}(\frac{q_{\psi}}{q_{\phi}}-1)^2] \\
    & =  \frac{1}{2}\Bigg[ \frac{\det(\Sigma_1)}{\det(\Sigma_2)} \exp\big(-\Tr(\Sigma_2^{-1}\Sigma_1) + \underbrace{d}_{=\Tr(\mathbf{I})} - \mu_1^T\Sigma_1^{-1}\mu_1 - \mu_2^T\Sigma_2^{-1}\mu_2 + 2\mu_1^T\Sigma_2^{-1}\mu_2 \big) \\ 
    & - 2 \frac{\det(\Sigma_1)^{\frac{1}{2}}}{\det(\Sigma_2)^{\frac{1}{2}}} \exp\big(-\frac{1}{2}\Tr(\Sigma_2^{-1}\Sigma_1) + \frac{d}{2}  -\frac{1}{2}\mu_1^T\Sigma_1^{-1}\mu_1 - \frac{1}{2}\mu_2^T\Sigma_2^{-1}\mu_2 + \mu_1^T\Sigma_2^{-1}\mu_2 \big) + 1\Bigg] . 
\end{align*}
When $\Sigma_1 = \Sigma_2 = \Sigma$, we have the following simplified expression 
\begin{align*}
    D_{\chi^2}(q_{\psi}||q_{\phi}) & =  \frac{1}{2}\Big[\exp\big(-||\mu_1 - \mu_2||_{\Sigma^{-1}}^2\big) -2 \exp\big(-\frac{1}{2}||\mu_1 - \mu_2||_{\Sigma^{-1}}^2\big)+ 1\Big] \\
    & = \frac{1}{2}[e^{-2s} - 2 e^{-s} + 1 ] = \frac{1}{2}(e^{-s}-1)^2 
\end{align*}
where $s = \frac{1}{2}||\mu_1 -\mu_2||_{\Sigma^{-1}}^2$. Letting $\frac{1}{2}(e^{-s}-1)^2 \leq b$ indicates $ -\sqrt{2b} \leq (e^{-s}-1) \leq \sqrt{2b}$. Since the value of $s$ is always non negative as a norm, $s\geq 0 \implies e^{-s} - 1 \leq 0 $. Thus, we have $ s \leq -\log\big((1-\sqrt{2b})_{+}\big)  $. Therefore, when the divergence constraint $D_{\chi^2}(q_{\psi}||q_{\phi}) \leq b$ is satisfied, we have $ ||\mu_1 -\mu_2||_{\Sigma^{-1}}^2 \leq -2\log\big((1-\sqrt{2b})_{+}\big)$, which is similar to \eqrefp{supp:eq:D_KL:two:gaussian} with a new constant for the distortion budget.

We make use of these relationships in our implementation as follows. We define $\mu$ to be functions that split the last layer of the output of the encoder part of the pretrained VAE, $g_{\theta_e}(x)$, and take the first half as the mean of the latent distribution. We let $\sigma$ be a function that takes the second half portion to be the diagonal values of the variance of the latent distribution. Then, our implementation of  $||\mu\big(g_{\theta_e}(x)\big) - \mu\big(g_{\theta_f} (g_{\theta_e}(x), w, y)\big)  ||_{\sigma(g_{\theta_e}(x)) ^{-1}}^2 \leq b$ is equivalent to $ ||\mu_1 -\mu_2||_{\Sigma^{-1}}^2 \leq b $. As shown in the previous two derivations, optimizing over this constraint is similar to optimizing over the defined KL and $\chi^2$-divergence constraints.
 
%
\subsection{Comparison and connection to differential privacy}
\label{supp:tech:approach:diff:privacy}
The basic intuition behind differential privacy is that it is very hard to tell whether the released sample $\tilde{z}$ originated from raw sample $x$ or $x'$ (or $z$ vs. $z^{\prime}$), thus, protecting the privacy of the raw samples. Designing such a releasing scheme, which is also often called a channel or mechanism, usually requires some randomized response or noise perturbation. Although the goal does not necessarily involve reducing the correlation between released data and sensitive labels, it is worth investigating the notion of differential privacy and comparing the performance of a typical scheme to our setting because of its prevalence in privacy literature. Furthermore, we establish how our approach can be \renyi~differentially private with certain specifications. In this section, we use the words channel, scheme, mechanism, and filter interchangeably as they have the same meaning in our context. Also, we overload the notation of $\varepsilon$ and $\delta$ because they are the conventions in differential privacy literature. 

\begin{definition}{$\big[ (\varepsilon, \delta)$-differential privacy \citep{dwork2014algorithmic} $\big]$ }
Let $\varepsilon, \delta \geq 0$. A channel $Q$ from space $\mathcal{X}$ to output space $\mathcal{Z}$ is differentially private if for all measurable sets $S \subset \mathcal{Z} $ and all neighboring samples $\{x_1, \hdots, x_n \}=x_{1:n} \in \mathcal{X} $ and $\{x_1', \hdots, x_n' \}=x_{1:n}' \in \mathcal{X} $, 
\begin{align}
    Q(Z\in S | x_{1:n}) \leq e^{\varepsilon} Q(Z\in S | x_{1:n}') + \delta .
\end{align}
\end{definition}
An alternative way to interpret this definition is that with high probability $1-\delta$, we have the bounded likelihood ratio $e^{-\varepsilon}\leq \frac{Q(z|x_{1:n})}{Q(z|x_{1:n}')} \leq e^{\varepsilon} $ (The likelihood ratio is close to 1 as $\varepsilon$ goes to $0$)\footnote{Alternatively, we may express it as the probability $ P (|\log \frac{Q(z|x_{1:n})}{Q(z|x_{1:n}')}| > \varepsilon) \leq \delta$}. Consequently, it is difficult to tell if the observation $z$ is from $x$ or $x'$ if the ratio is close to 1. In the following discussion, we consider the classical Gaussian mechanism $\tilde{z} = f(z) + w$,  where $f$ is some function (or query) that is defined on the latent space and $w \sim \mathcal{N}(0, \sigma^2\mathbf{I})$.  We first include a theorem from \cite{dwork2014algorithmic} to disclose how differential privacy using the Gaussian mechanism can be satisfied by our baseline implementation and constraints in the robust optimization formulation. We denote a pair of neighboring inputs as $z\simeq z'$ for abbreviation. 

\begin{theorem}{$\big[$\cite{dwork2014algorithmic} Theorem A.1$\big]$}
\label{supp:thm:gaussian:mech:dp}
For any $\varepsilon, \delta \in (0, 1)$, the Gaussian mechanism with parameter $\sigma \geq \frac{L \sqrt{2\log(\frac{1.25}{\delta})}}{\varepsilon}$ is $(\varepsilon, \delta)$-differential private, where $L = \max_{z \simeq z'}||f(z) - f(z')||_2$ denotes the $l_2$-sensitivity of $f$. 
\end{theorem}
Next, we introduce a relaxation of differential privacy that is based on the R$\acute{\text{e}}$nyi divergence.

\begin{definition}
$\big[$\renyi~divergence (\cite{mironov2017renyi}, Definition 3)$\big]$. Let $P$ and $Q$ be distributions on a space $\mathcal{X}$ with densities $p$ and $q$ (with respect to a measure $\mu$). For $\alpha \in [1, \infty]$, the \renyi-$\alpha$-divergence between $P$ and $Q$ is 
\begin{equation}
    D_{\alpha}(P||Q) = \frac{1}{\alpha-1} \log \int \Big(\frac{p(x)}{q(x)}\Big)^{\alpha} q(x) d\mu(x),
\end{equation}
where the values $\alpha \in \{1, \infty\}$ are defined in terms of their respective limits. 
\end{definition}

In particular, $\lim_{\alpha \downarrow 1}D_{\alpha}(P||Q) = D_{\text{KL}}(P||Q)$. We use \renyi~divergences because they satisfy $\exp\big((\alpha-1) D_{\alpha}(P||Q)\big) = D_f(P||Q)$ when $f$-divergence is defined by $f(t)=t^{\alpha}$. And such an equivalent relationship has a natural connection with the $f$-divergence constraint in our robust optimization formulation. With this definition, we introduce the \renyi~differential privacy, which is a strictly stronger relaxation than the $(\varepsilon, \delta)$-differential privacy relaxation. 

\begin{definition} \label{supp:renyi_priv}
$\big[$ \renyi~differential privacy $\big($\cite{mironov2017renyi}, Definition 4$\big) \big]$. Let $\varepsilon \geq 0$ and $\alpha \in [1, \infty]$. A mechanism $F$ from $\mathcal{X}^n$ to output space $\mathcal{Z}$ is $(\varepsilon, \alpha)$- \renyi~private if for all neighboring samples $x_{1:n}, x_{1:n}^{\prime} \in \mathcal{X}^n$,
\begin{equation}
    D_{\alpha}\big(F(\cdot | x_{1:n})||F(\cdot | x_{1:n}^{\prime}) \big) \leq \varepsilon.
\end{equation}
\end{definition}

For the basic Gaussian mechanism, we apply the additive Gaussian noise on $z$ directly to yield 
\begin{align}
    \tilde{z} = z + w, \quad w\sim \mathcal{N}(0, \sigma^2 I) \label{app:gaussian:dp:basic}. 
\end{align}

We first revisit the basic Gaussian mechanism and its connection to \renyi-differential privacy. 

\begin{theorem}
Let $L = \max_{z, z'\in \mathcal{Z}}||z-z'||_2, \forall z, z'$ and $w \sim \mathcal{N}(0, \sigma^2I)$. Then the basic Gaussian mechanism shown in \eqrefp{app:gaussian:dp:basic} is $(\varepsilon, \alpha)-$\renyi \text{ } private if $\sigma^2 = \frac{\alpha L^2}{2\varepsilon}$.
\label{thm:basic:add:gaussian:dp}
\end{theorem}

\emph{Proof of \Thmref{thm:basic:add:gaussian:dp}}. Considering two examples $z$ and $z'$, we calculate the \renyi~divergence between $(z+w) \sim \mathcal{N}(z, \sigma^2 I)$ and $(z'+ w) \sim \mathcal{N}(z', \sigma^2 I)$:
\begin{align}
    D_{\alpha}(\mathcal{N}(z, \sigma^2I) || \mathcal{N}(z', \sigma^2 I ) ) \explainEq{i} \frac{\alpha}{2 \sigma^2} ||z -z'||_2^2 \leq  \frac{\alpha}{2 \sigma^2} \max_{z, z'\in \mathcal{Z}}||z-z'||_2^2 = \frac{\alpha L^2}{2 \sigma^2} = \varepsilon.
\end{align}
The equality (i) is shown in the deferred proofs in \eqrefp{app:deferred:renyi:two:gaussian:sameVar}. Arranging $\sigma^2 = \frac{\alpha L^2}{2\varepsilon}$ gives the desired result. Although this result is already known as Lemma 2.5 in \citep{bun2016concentrated}, we simplify the proof technique.   

Because normal distributions are often used to approximate data distributions in real applications, we now consider the scenario where the original data $z$ is distributed as a multivariate normal $\mathcal{N}(\mu, \Sigma)$. This is a variant of additive Gaussian mechanism, as we add noise to the mean, i.e. 
\begin{align}
    \tilde{\mu} = \mu + \sigma w, \quad w \sim \mathcal{N}(0, I).
\end{align}
We have the following proposition. 
\begin{proposition}\label{appx:prop:1}
Suppose a dataset $Z$ with sample $z$ is normally distributed with parameters $\mathcal{N}(\mu, \Sigma)$, and there exists a scalar $\sigma^2$ such that $\max_{z, z^{\prime} \in Z } ||z - z^{\prime}|| \leq \sigma^4\Tr(\Sigma^{-1}) $. Under the simple additive Gaussian mechanism, when both $\mathbb{E}\big[D_{\alpha}\big(\mathcal{N}(\mu+\sigma w, \Sigma) || \mathcal{N}(\mu, \Sigma) \big)\big] \leq b $, then such a mechanism satisfies $(b, \alpha)$ \renyi~differential privacy.   
\end{proposition} 

\emph{Proof of \Propref{appx:prop:1}}:
\begin{align}
    b & \explainGeq{\text{use assump.}} \mathbb{E}\Big[ D_{\alpha}\big(\mathcal{N}(\mu+\sigma w, \Sigma) || \mathcal{N}(\mu, \Sigma) \big)\Big] \\
    & = \mathbb{E}\Big[\frac{\alpha}{2} (\mu+\sigma w -\mu)^T\Sigma^{-1}(\mu + \sigma w - \mu) \Big] = \frac{\alpha}{2}\sigma^2\mathbb{E}[w^T\Sigma^{-1}w]\\
    & = \frac{\alpha \sigma^2}{2}\mathbb{E}[\Tr(\Sigma^{-1}ww^T)] = \frac{\alpha\sigma^2}{2}\Tr(\mathbb{E}[\Sigma^{-1}ww^T]) \\
    & = \frac{\alpha\sigma^2}{2} \Tr\Big(\E[\Sigma^{-1}]\underbrace{\E[ww^T]}_{I}\Big) = \frac{\alpha\sigma^2}{2} \Tr(\Sigma^{-1}) \\
    & \explainGeq{\text{use assump.}} \frac{\alpha\sigma^2}{2}\frac{1}{\sigma^4} \max_{z, z' \in  {Z}}||z-z'|| = \frac{\alpha}{2\sigma^2}\max_{z, z' \in  {Z}}||z-z'|| \geq D_{\alpha}(\mathcal{N}(z, \Sigma) || \mathcal{N}(z', \Sigma)) \label{prop:revise1:end}
\end{align}

With a prescribed budget $b$, the predetermined divergence $\alpha$, and a known data covariance $\Sigma$, we can learn an adjustable $\sigma$. Moreover, if $\sigma^4 \geq \frac{\max||z - z^{\prime}||}{\Tr(\Sigma^{-1})}$ and $b\geq \frac{\alpha\sigma^2}{2}\Tr(\Sigma^{-1})$, we have $(b, \alpha)$ \renyi~differential privacy.



To build the connection between the \renyi~differential privacy and our constrained robust optimization, we explicitly impose some matrix properties of our linear filter. We denote the matrix $\Gamma$ to be the linear filter, the one-hot vector $y_s$ to represent private label $y$, and $w$ to be standard Gaussian noise. This method can be considered as a linear filter mechanism, a variant of additive transformed Gaussian mechanism. The output $\tilde{z}$ generated from $z$ and $w$ can be described as  
\begin{equation}
    \tilde{z} = z + \Gamma \begin{bmatrix}
    w \\
    y_s
    \end{bmatrix} = z + \begin{bmatrix}
    A& V\end{bmatrix}
    \begin{bmatrix}
    w \\
    y_s
    \end{bmatrix} = z + Aw + Vy_s.
\end{equation}
We decompose the matrix $\Gamma$ into two parts: $A$ controls the generative noise and $V$ controls the bias. 
Now we present the following theorem. 

\begin{theorem} \label{supp:thm:filter_priv}
Let matrix $\Gamma$ be decomposed into $\Gamma= \begin{bmatrix} A & V
\end{bmatrix}$, $\sigma_{min} > 0$ be the minimum eigenvalue of $AA^T$,  $||V||_1 = \tau$, and $L=\max_{z,z'\in \mathcal{Z}} ||z - z'||_2$, then the mechanism 
\begin{equation}
    \tilde{z} = z_{1:n} + Aw + Vy_{s\{1:n\}} \label{app:eq:mech:renyi:dp}
\end{equation}
satisfies $(\varepsilon, \alpha)$-\renyi~privacy, where $\varepsilon=\frac{\alpha d}{2\sigma_{min}}(L^2+2\tau^2)$ and $z_{1:n}, y_{s\{1:n\}}$ are $d$-dimensional samples. 
\label{app:theo:2}
\end{theorem}

\emph{Proof of \Thmref{app:theo:2}}.
Consider two examples $(z, y_s)$ and $(z', y_s')$. Because $w \sim \mathcal{N}(0, I)$, the corresponding output distributions yielded from these two examples are $\mathcal{N}(z + Vy_s, AA^T)$ and $\mathcal{N}(z' + Vy_s', AA^T)$ through the linear transformation of Gaussian random vector $w$. The resulting \renyi \text{} divergence is 
\begin{subequations}
\begin{align}
    & D_{\alpha}\big(\mathcal{N}(z + Vy_s, AA^T) || \mathcal{N}(z' + Vy_s', AA^T) \big) \\
    & \explainEq{i} \frac{\alpha}{2}(z - z' + Vy_s - Vy_s')^T (AA^T)^{-1}(z - z' + Vy_s - Vy_s') \label{app:renyi:eq:2} \\
    & \explainLeq{ii} \frac{\alpha d}{2\sigma_{min}}||z - z' + Vy_s - Vy_s'||_2^2 \label{app:renyi:eq:3} \\
    & \explainLeq{iii}
    \frac{\alpha d}{2\sigma_{min}}\big{\|}z - z' +  2|v_{j^*}| \big{\|}_2^2 \label{app:renyi:eq:4}  \\
    & \explainLeq{iv} \frac{\alpha d}{2\sigma_{min}}(||z - z'||_2^2 + 2||V||_1^2 ) \label{app:renyi:eq:5} \\
    & \leq \frac{\alpha d}{2\sigma_{min}}(\max_{z,z\in \mathcal{Z}}||z-z'||_2^2 + 2||V||_1^2)
\end{align}
\end{subequations}
where equality (i) uses the property of \renyi~divergence between two Gaussian distributions \citep{gil2013renyi} that is provided in the deferred proof of  \eqrefp{app:deferred:renyi:two:gaussian:sameVar}, and inequality (ii) uses the assumption that $\sigma_{min}$ is the minimum eigenvalue of $AA^T$ (so that $AA^T \geq \sigma_{min}I$). Because $y_s$ is a one-hot vector, $Vy_s$ returns a particular column of $V$ where the column index is aligned with the index of non-zero entry  of $y_s$. Inequality (iii) picks the index $j^*$ from column vectors $[v_1,...v_j,...,v_K]= V$ (if the label y has $K$ classes) such that $j^* = \argmax_{j} \sum_i |v_{ij}| $. We denote $|v_{j^*}| $ as a vector taking absolute value of each entry in $v_{j^*}$. The final (iv) simply applies the triangle inequality and $\max_{j} \sum_i |v_{ij}| = ||V||_1$ (maximum absolute column sum of the matrix).  The remainder of the proof is straightforward by changing $||z-z'||_2^2$ to $\max_{z,z\in \mathcal{Z}}||z-z'||_2^2$ in  \eqrefp{app:renyi:eq:5} as the mechanism runs through all samples. Finally, setting  
$\frac{\alpha d}{2\sigma_{min}}(\max||z - z'||_2^2 + 2||V||_1^2 ) = \varepsilon$ with substitutions of $L$ and $\tau$ gives the desired result.\\~\\
Consequently, we connect the $\alpha$-\renyi~divergence to differential privacy by presenting Corollary \ref{app:coroll:rdp:dp}.   
\begin{corollary}
 The $\big(\frac{\alpha d}{2\sigma_{min}}(L^2+2\tau^2), \alpha \big)$ \renyi~differential privacy mechanism also satisfies $\big( \frac{\alpha d}{2\sigma_{min}}(L^2+2\tau^2)  + \frac{1}{\alpha-1}\log \frac{1}{\delta}, \delta\big)$ differential privacy for any $\delta > 0$. \label{app:coroll:rdp:dp}
\end{corollary}
\emph{Proof of \Corollref{app:coroll:rdp:dp}}. The proof is an immediate result of proposition 3 in Minronov's work \citep{mironov2017renyi} when we set $\varepsilon = \frac{\alpha d}{2\sigma_{min}}(L^2+2\tau^2)$. \\~\\
\emph{Remark:} When having high privacy with small $\varepsilon$, we set $\sigma_{min}$ to be large. This can be seen from the following relationship.
\begin{subequations}
\begin{align}
    & D_f = \exp\big( (\alpha -1) D_{\alpha} \big)
    \explainLeq{\thmref{app:theo:2}}\exp\big( (\alpha -1) \frac{\alpha d(L^2 + 2\tau^2)}{2\sigma_{min}} \big) \explainLeq{def} b \\
    & \implies \sigma_{min} \geq \Big[ \frac{\alpha(\alpha -1)d(L^2 + 2\tau^2)}{2\log b}\Big]_{+} . 
\end{align}
\end{subequations}
One intuitive relationship is that if $L$ is large (i.e. $\max_{z,z' \in \mathcal{Z}}||z - z'||_2$ is large), the $\sigma_{min}$ goes up with a quadratic growth rate with respect to $L$. Because large $L$ indicates $z$ is significantly distinct from $z'$, which requires a large variance of the noise (i.e. large $AA^T$) to obfuscate the original samples, $\sigma_{min}$ also shrinks with logarithmic growth of $b$ when the privacy level is less stringent.\\~\\ However we also notice that the eigenvalues of $AA^T$ cannot grow to infinity, as shown in \Propref{app:thm:prop:AA_T:bound}.


\begin{proposition}\label{app:thm:prop:AA_T:bound}
Suppose a dataset $\mathcal{Z}$ with sample z, a $d$-dimensional vector, is normally distributed with parameters $\mathcal{N}(\mu, \Sigma)$, and there exists $A$ and $V$ such that $\Tr(\Sigma^{-1}AA^T) \geq \frac{d}{\sigma_{\min}} \Big( \max||z-z' ||_2^2 + 2 ||V||_1^2 \Big)$ where $\sigma_{\min}$ is the minimum singular value of $AA^T$. Under our linear filter mechanism, when $\mathbb{E}\Big[{D}_{\alpha}\big(\mathcal{N}(\mu + Vy_s + Aw, \Sigma) ||\mathcal{N}(\mu, \Sigma)\big) \Big] \leq b$, then such a mechanism satisfies $(b, \alpha)$ \renyi~differential privacy.  
\end{proposition}

\emph{Proof of \Propref{app:thm:prop:AA_T:bound}}. 

\begin{subequations}
\begin{align}
    b \explainGeq{\text{use assump.}} & \mathbb{E}\Big[{D}_{\alpha}\Big\{ \mathcal{N}(\mu + Vy_s + Aw, \Sigma) || \mathcal{N}(\mu, \Sigma) \Big\}\Big] = \frac{\alpha}{2} \E \Big[ (Vy_s+Aw)^T \Sigma^{-1}(Vy_s+Aw) \Big] \\
    & = \frac{\alpha}{2} \E \Big[ \Tr\big(\Sigma^{-1} (Vy_s + Aw)(Vy_s + Aw)^T \big) \Big] \\
    & = \frac{\alpha}{2} \E\Big[ \Tr \big(\Sigma^{-1}( Vy_s y_s^TV^T + Vy_s w^TA^T + Awy_s^TV^T + Aww^TA^T) \big) \Big]   \\
    & = \frac{\alpha}{2} \Tr\Big[ \E\big(\Sigma^{-1} (Vy_sy_s^TV^T + Aww^TA^T)\big) \Big] \\
    & = \frac{\alpha}{2} \Tr\Big[ \E\big(\Sigma^{-1} (Vy_sy_s^TV^T)\big) + \Sigma^{-1} A\E[ww^T]A^T  \Big] \\
    & = \frac{\alpha}{2} \Tr\Big[ \E\big(\Sigma^{-1} (Vy_sy_s^TV^T)\big) + \Sigma^{-1} AA^T  \Big] \\
    & = \frac{\alpha}{2} \Big( \E[  \Tr \underbrace{(y_s^TV^T\Sigma^{-1}Vy_s)}_{>0, \text{as positive definite}} ] + \Tr(\Sigma^{-1}AA^T) \Big) \\
    & \geq \frac{\alpha}{2} \Tr(\Sigma^{-1}AA^T) 
    \explainGeq{\text{use assump.}} \frac{\alpha d}{2\sigma_{\min}} (\max||z-z'||^2 + 2||V||_1^2) \explainGeq{\text{use theorem 3}} D_{\alpha}(f(z)||f(z')) 
\end{align}
\end{subequations}

The $\sigma_{\min}$ is the minimum singular value of $AA^T$ and $f$ is some function mapping.  

If $\Sigma$ is a diagonal matrix, we want to find such an $A$ that holds the following property that 
\begin{align}
    b \geq \frac{\alpha}{2}\Tr(\Sigma^{-1}AA^T) \geq \frac{\alpha d}{2\sigma_{(\Sigma, \max)}}\Tr(AA^T) = \frac{\alpha d}{2\sigma_{(\Sigma, \max)}}||A||_{F}, 
\end{align}
where $\sigma_{(\Sigma, \max)}$ is the max singular value of $\Sigma$. Thus, the eigenvalues of $AA^T$ is upper bounded.  
Therefore, with a prescribed $b$, a predetermined $\alpha$, and a known empirical covariance $\Sigma$, we can learn a linear filter $\Gamma=[A, V]$. Moreover, if $A$ and $V$ satisfies $\frac{\alpha d}{2\sigma_{(\Sigma, \max)}}||A||_{F} \leq b$ and $\Tr(\Sigma^{-1}AA^T) \geq \frac{d}{\sigma_{\min}}\big( \max_{z,z^{\prime}\in Z}||z-z^{\prime}||^2 + 2||V||_1^2\big)$, then we have $(b, \alpha)$ \renyi~differential privacy.

\subsection{Why does cross-entropy loss work}
In this section, we explain the connection between cross-entropy loss and mutual information to give intuition for why maximizing the cross-entropy loss in our optimization reduces the correlation between released data and sensitive labels. Given that the encoder part of the VAE is fixed, we focus on the latent variable $z$ for the remaining discussion in this section.   

The mutual information between latent variable $z$ and sensitive label $y$ can be expressed as follows  
\begin{align}
    I(z;y) & =\mathbb{E}_{q(z, y)}\Big[\log q(y|z) - \log q(y) \Big] \\
    & = \mathbb{E}_{q(z, y)}\big[ \log q(y|z) - \log p(y|z) - \log q(y) + \log p(y|z) \big] \\
    & = \mathbb{E}_{q(y|z)q(z)}\big[ \log \frac{q(y|z)}{p(y|z)} \big] + \mathbb{E}_{q(z,y)} [\log p(y|z) ] - \mathbb{E}_{q(y|z)q(y)}[\log q(y)] \\
    & = \underbrace{\mathbb{E}_{q(z)}\Big[ D_{KL}\big(q(y|z)|| p(y|z)\big) \Big]}_{\geq 0} + \mathbb{E}_{q(z, y)}[\log p(y|z)] + H(y) \\
    & \geq \mathbb{E}_{q(z, y)}[\log p(y|z)] + H(y), \label{supp:eq:I:vari:low:bd}
\end{align}
where $q$ is the data distribution, and $p$ is the approximated distribution, which is similar to the last logit layer of a neural network classifier. Then, the term $-\mathbb{E}_{q(z, y)}[\log p(y|z)]$ is the cross-entropy loss $H(q, p) $ (the corresponding negative log-likelihood is $-\mathbb{E}_{q(z, y)}[\log p(y|z)]$ ). In classification problems, minimizing the cross-entropy loss enlarges the value of $\mathbb{E}_{q(z, y)}[\log p(y|z)]$. Consequently, this pushes the lower bound of $I(z; y)$ in \eqrefp{supp:eq:I:vari:low:bd} as high as possible, indicating high mutual information. 

However, in our robust optimization, we maximize the cross-entropy, thus, decreasing the value of $\mathbb{E}_{q(z, y)}[\log p(y|z)]$ (more specifically, it is $\mathbb{E}_{q(\tilde{z}, y)}[\log p(y|\tilde{z})]$, given the mutual information we care about is between the new representation $\tilde{z}$ and sensitive label $y$ in our application). Thus, the bound of \eqrefp{supp:eq:I:vari:low:bd} has a lower value, which indicates the mutual information $I(\tilde{z}; y)$ can be lower than before. Such observations can also be supported by the empirical results of mutual information shown in \figref{fig:mnist:mi:plot:case1:case2}.  
%

\subsection{Dependency of filter properties and tail bounds}
\label{supp:class_acc}
In terms of classifying private attributes, we claim that new perturbed samples $\tilde{z}$ become harder to classify correctly compared to the original samples $z$. To show this, we further simplify the setting by considering a binary classification case with data $(z, y) \in \mathbb{R}^d \times \{\pm 1\}$. We consider linear classifiers $\theta^Tz$ using zero-one loss based on the margin penalty $\xi > 0$. More precisely, we define the loss $\ell_{\xi}(\theta, (z, y)) = \mathbbm{1}\{ \theta^Tz y \leq \xi \}$. Then the expected loss
\begin{align}
    L_{\xi}(\theta; z) = \mathbb{E}[\ell_{\xi}(\theta, (z,y) )]= P(\theta^T zy \leq \xi).
\end{align}
Recall that $\tilde{z} = z + Aw + Vy_s$, where $y_s$ is the one-hot vector that represents $y$. We have the following expressions:  
\begin{align}
    P(\theta^Tzy \leq \xi) & = P\big(\theta^T(\tilde{z} - Aw - Vy_s)y \leq \xi \big) \\
    & = P\big(\theta^T\tilde{z}y -y\theta^TAw - y\theta^TVy_s \leq \xi \big) \\
    & \explainLeq{i} P\big(\{ \theta^T\tilde{z}y \leq \xi \} \cap \big\{\min\{ - \theta^TAw - \theta^Tv_1, \theta^TAw + \theta^Tv_2 \} \leq 0 \big\}  \big) \\
    & \explainEq{ii} P(\theta^T\tilde{z}y \leq \xi) P (\min\{ - \theta^TAw - \theta^Tv_1, \theta^TAw + \theta^Tv_2 \} \leq 0) \\
    & \leq P(\theta^T\tilde{z}y \leq \xi) \max \big\{ P( \theta^TAw \geq -\theta^Tv_1), P( \theta^TAw \leq -\theta^Tv_2) \big\} \\
    & \explainLeq{iii}  P(\theta^T\tilde{z}y \leq \xi) \underbrace{\exp \Big(-\frac{||\theta||_2^2 \tau^2}{2(\theta^TAA^T\theta)}\Big)}_{<1}. \label{app:eq:derive:margin:classify:loss}
\end{align}

The inequality (i) uses the fact that the matrix $V$ multiplied by the one-hot vector $y_s$ returns the column vector with index aligned with the non-zero entry in $y_s$. We explicitly write out column vector $v_1, v_2$ in this binary classification setting when $y=\pm 1$. The equality (ii) uses the fact that $w$ is the independent noise when we express out $y$. 
The inequality (iii) uses the concentration inequality of sub-Gaussian random variables [\cite{rigollet2015high}, Lemma 1.3]. Specifically, since $\E(\theta^TAw)=0$, we have $P(\theta^TAw ) \leq \exp\big(- \frac{(-\theta^Tv_1)^2}{2\theta^TAA^T\theta }\big)$ and $P(\theta^TAw ) \leq \exp\big(- \frac{(\theta^Tv_2)^2}{2\theta^TAA^T\theta }\big)$. We then apply the Cauchy-Schwarz inequality on $\theta^T v_1$ (also on $\theta^T v_2$) and let $\tau$ be the maximum $l_2$-norm of $v_1$ and $v_2$, i.e. $\tau=\max\{||v_1||_2, ||v_2||_2\}$. Therefore, by rearranging \eqrefp{app:eq:derive:margin:classify:loss} we show that  
\begin{align}
    P(\theta^T\tilde{z}y \leq \xi ) \geq P(\theta^Tzy \leq \xi ) \bigg( \exp \Big(-\frac{||\theta||_2^2 \tau^2}{2(\theta^TAA^T\theta)}\Big)\bigg)^{-1}
    \implies L_{\xi}(\theta; \tilde{z}) \geq L_{\xi}(\theta; z),
\end{align} 
which indicates that classifying $\tilde{z}$ is harder than classifying $z$ under 0-1 loss with the margin based penalty.\\ 
\emph{Remark:} Inequality \eqrefp{app:eq:derive:margin:classify:loss} provides an interesting insight that a large Frobenius norm of $A$, i.e. $||A||_{\text{F}}$, gives higher loss on classifying new samples. To see why it holds, we apply the trick $\theta^TAA^T\theta = \Tr(\theta^TAA^T\theta ) = \Tr(AA^T\theta\theta^T)=||A||_{\text{F}}^2 \Tr(\theta\theta^T) $. Thus a large value of $||A||_{\text{F}}^2$ pushes $\exp\Big(-\frac{||\theta||_2^2 ||v^*||_2^2}{2(\theta^TAA^T\theta)}\Big)$ to small values, which increases the $L_{\xi}(\theta; \tilde{z})$.\\
As mentioned in \cite{nguyen2010estimating} and \cite{duchi2018multiclass}, other convex decreasing loss functions that capture margin penalty can be surrogates of 0-1 loss, e.g. the hinge loss $L(t)= (1-t)_{+} $ or logistic loss $L(t)= \log(1+\exp(-t))$ where $t=\theta^Tzy$.


\subsection{Experiment Details}
%
All experiments were performed on Nvidia GTX 1070 8GB GPU with Python 3.7 and Pytorch 1.2.

\label{supp:nn:archi:explain}
\subsubsection{VAE architecture}
The MNIST dataset contains 60000 samples of gray-scale handwritten digits with size 28-by-28 pixels in the training set, and 10000 samples in the testing set.  
When running experiments on MNIST, we convert the 28-by-28 images into 784 dimensional vectors and construct a network with the following structure for the VAE:
\begin{align*}
    & x \rightarrow \text{FC}(300) \rightarrow \text{ELU} \rightarrow \text{FC}(300) \rightarrow \text{ELU} \rightarrow \text{FC}(20) (\text{split $\mu$ and $\Sigma$ to approximate $q(z|x)$}) \\
    & z \rightarrow \text{FC}(300) \rightarrow \text{ELU} \rightarrow \text{FC}(300) \rightarrow \text{ELU} \rightarrow \text{FC}(784).
\end{align*}
The UCI-Adult data contains 48842 samples with 14 attributes. Because many attributes are categorical, we convert them into a one-hot encoding version of the input. We train a VAE with the latent dimension of 10 (both mean and variance with dimension of 10 for each) with two FC layers and ELU activation functions for both encoder and decoder.  

The UCI-abalone data has 4177 samples with 9 attributes. We pick sex and ring as private and utility labels, leaving 7 attributes to compress down. The VAE is just single linear layer with 4 dimensional output of embedding, having 4-dimensional mean and variance accordingly.   

The aligned CelebA dataset contains 202599 samples. We crop each image down to 64-by-64 pixels with 3 color (RGB) channels and pick the first 182000 samples as the training set and leave the remainder as the testing set. The encoder and decoder architecture for CelebA experiments are described in Table~\ref{tab:encode:archi} and Table~\ref{tab:decode:archi}.

\begin{table}[!hpbt]
\vspace{-0.1in}
    \caption{Encoder Architecture in CelebA experiments. We use the DenseNet \citep{huang2017densely} architecture with a slight modification to embed the raw images into a compact latent vector, with growth rate $m=12$ and depth $k=82$}
    \label{tab:encode:archi}
    \centering
    \begin{small}
    \begin{tabular}{c|c|c}
    \toprule
         \textbf{Name} & \textbf{Configuration} & \textbf{ Replication} \\
    \midrule
         \multirow{2}{*}{initial layer} & conv2d=(3, 3), stride=(1, 1), & \multirow{2}{*}{1} \\
         & padding=(1, 1), channel in = 3, channel out = $2m$ &  \\
    \midrule 
         \multirow{3}{*}{dense block1} & batch norm, relu, conv2d=(1, 1), stride=(1, 1), & \\
         & batch norm, relu, conv2d=(3, 3), stride=(1, 1), & 12 \\
         & growth rate = $m$, channel in = 2$m$  & \\ 
    \midrule      
        \multirow{4}{*}{transition block1} &  batch norm, relu, & \multirow{4}{*}{1} \\
        & conv2d=(1, 1), stride=(1, 1), average pooling=(2, 2), &  \\
        & channel in = $ \frac{(k-4)}{6} m + 2m$, & \\
        & channel out = $\frac{(k-4)}{12} m + m$ &  \\
    \midrule 
    \multirow{3}{*}{dense block2} & batch norm, relu, conv2d=(1, 1), stride=(1, 1), & \multirow{3}{*}{12} \\
         & batch norm, relu, conv2d=(3, 3), stride=(1, 1), & \\
         & growth rate=$m$, channel in = $ \frac{(k-4)}{12} m + m$, & \\
    \midrule      
        \multirow{4}{*}{transition block2} &  batch norm, relu, & \multirow{4}{*}{1}\\
        & conv2d=(1, 1), stride=(1, 1), average pooling=(2, 2), & \\
        & channel in = $\frac{(k-4)}{6} m  + \frac{(k-4)}{12} m + m $ &  \\ 
        & channel out= $\frac{1}{2}\big(\frac{(k-4)}{6} m  + \frac{(k-4)}{12} m + m \big)$ & \\
    \midrule 
    \multirow{3}{*}{dense block3} & batch norm, relu, conv2d=(1, 1), stride=(1, 1), & \multirow{3}{*}{ 12} \\
         & batch norm, relu, conv2d=(3, 3), stride=(1, 1), & \\ 
         & growth rate = $m$, channel in = $\frac{1}{2}\big(\frac{(k-4)}{6} m  + \frac{(k-4)}{12} m + m \big)$ & \\ 
    \midrule      
        \multirow{4}{*}{transition block3} &  batch norm, relu, & \multirow{4}{*}{1} \\ 
        & conv2d=(1, 1), stride=(1, 1), average pooling=(2, 2), \\
        & channel in = $\frac{1}{2}\big(\frac{(k-4)}{6} m  + \frac{(k-4)}{12} m + m \big) +\frac{(k-4)}{6}m $  &  \\
        & channel out = $\frac{1}{2}\big(\frac{1}{2}\big(\frac{(k-4)}{6} m  + \frac{(k-4)}{12} m + m \big) +\frac{(k-4)}{6}m \big)$ & \\
    \midrule 
        output layer & batch norm, fully connected 100  & 1 \\
    \bottomrule     
    \end{tabular}
    \end{small}
\end{table}

\begin{table}[!hpbt]
\vspace{-0.0in}
    \caption{Decoder Architecture in CelebA experiments. }
    \label{tab:decode:archi}
    \centering
    \begin{small}
    \begin{tabular}{c|c|c}
    \toprule
    \textbf{Name} & \textbf{Configuration} & \textbf{ Replication} \\
    \midrule
    initial layer & fully connected 4096 & 1 \\
    \midrule
    reshape block & resize 4096 to $256 \times 4 \times 4$ & 1 \\
    \midrule 
    \multirow{3}{*}{deccode block} & conv transpose=(3, 3), stride=(2, 2), & \multirow{3}{*}{4}\\
    & padding=(1, 1), outpadding=(1, 1), & \\
    & relu, batch norm & \\
    \midrule 
    \multirow{2}{*}{decoder block} & conv transpose=(5, 5), stride=(1, 1), & \multirow{2}{*}{1}\\
    & padding=(2, 2) &  \\
    \midrule 
    \end{tabular}
    \end{small}
\end{table}

\subsubsection{Filter Architecture}
We use a generative linear filter throughout our experiments. In the MNIST experiments, we compressed the latent embedding down to a 10-dim vector. For \textbf{MNIST Case 1}, we use a 10-dim Gaussian random vector $w$ concatenated with a 10-dim one-hot vector $y_s$ representing digit id labels, where $w \sim \mathcal{N}(0, I)$ and $y_s \in \{0 ,1\}^{10}$. We use the linear filter $\Gamma$ to ingest the concatenated vector and add the corresponding output to the original embedding vector $z$ to yield $\tilde{z}$. Thus the mechanism is
\begin{align}
    \tilde{z}=f(z, w, y) = z + \Gamma \begin{bmatrix} w \\ y_s\end{bmatrix} \label{appx:filter:scheme},
\end{align}
where $\Gamma \in \mathbb{R}^{10\times 20}$ is a matrix. For \textbf{MNIST Case 2}, we use a similar procedure except the private label $y$ is a binary label (i.e. digit value $\geq 5$ or not). Thus, the corresponding one-hot vector is 2-dimensional. Since we keep $w$ to be a 10-dimensional vector, the corresponding linear filter $\Gamma$ is a matrix in $\mathbb{R}^{10 \times 12}$. 

In the experiment of CelebA, we create the generative filter following the same methodology in \eqrefp{appx:filter:scheme}, with some changes on the dimensions of $w$ and $\Gamma$ because images of CelebA are bigger than MNIST digits.\footnote{We use a VAE type architecture to compress the image down to a 100 dimensional vector, then enforce the first 50 dimensions as the mean and the second 50 dimensions as the variance} Specifically, we set $w \in R^{50}$ and $A \in \mathbb{R}^{50 \times 52}$.  

%
%
%
\subsubsection{Adversarial classifiers}
In the MNIST experiments, we use a small architecture consisting of neural networks with two fully connected layers and an exponential linear unit (ELU) to serve as the privacy classifiers, respectively. The specific structure of the classifier is depicted as follows:
\begin{align*}
    & z \text{ or } \tilde{z} \rightarrow \text{FC}(15) \rightarrow \text{ELU} \rightarrow \text{FC}(y) [\text{ or FC}(u) ].
\end{align*} 
We use linear classifier for UCI-adults and UCI-abalone with input dimension of 10 and 4 which aligns with the embedding dimensions respectively. 
In the CelebA experiments, we construct a two-layered neural network as follows:
\begin{align*}
    & z \text{ or } \tilde{z} \rightarrow \text{FC}(60) \rightarrow \text{ELU} \rightarrow \text{FC}(y) [\text{ or FC}(u) ].
\end{align*} 
The classifiers ingest the embedded vectors and output unnormalized logits for the private label or utility label. The classification results of CelebA can be found in Table~\ref{tab:celebA:clf:acc:results}. 

\subsubsection{Other hyper-parameters}
When we first train the VAE type models using loss function in \eqrefp{eq:supp:vae:loss:variant}, we pick multiple values such as $\gamma=\{0.01, 0.1, 1, 10, 100\}$, $\kappa = \{1, 10\}$, and $C_{L}=\{2, 4, 6\}$ to evaluate the performance. A combination of $\gamma=0.1, \kappa=1, C_L=4$ yields the smallest loss among all the options.

In the min-max training using the objective in \eqrefp{eq6:appx:obj}, we pick multiple betas ($\beta=\{0.1, 0.5, 1, 2, 4\}$) and report the results when $\beta=2$ (in MNIST and CelebA) and $\beta=1$ (in UCI-Adult and UCI-abalone) because this gives the largest margin between accuracy of utility label and accuracy of private label [e.g. highest ($acc_u - acc_y$)]. We also used the relaxed soft constraints mentioned in \eqrefp{eq6:ctr:relax} by setting $\lambda_1 = \lambda_2 = 1000$ and divide them in halves every 500 epochs with a minimum clip value of 2. We train 1000 epochs for MNIST, UCI-Adult, and UCI-abalone, and 10000 epochs for CelebA.    

We use Adam optimizer \citep{kingma2014adam} throughout the experiments with learning rate 0.001 and batch size 128 for MNIST, UCI-Adult and UCI-abalone. In CelebA experiment, the learning rate is 0.0002 and the batch size is 24.  


\subsubsection{Distributed training setting}
\label{supp:distributed_details}
This section provides a more detailed look into how our scheme can be run in a local and distributed fashion through an example experiment on the MNIST dataset with 2 independent users. The first user adopts the label of digit $\geq 5$ or not as private and odd or even as the utility label. The second user prefers the opposite and wishes to privatize odd or even and maintain $\geq 5$ or not as the utility label. We first partition the MNIST dataset into 10 equal parts where the first part belongs to one user and the second part belongs to the other user. The final eight parts have already been made suitable for public use either through privatization or because they do not contain information that their original owners have considered sensitive. Each part is then encoded into their 10-dimensional representations and passed onto the two users for the purpose of training an appropriate classifier rather than one trained on a single user's biased dataset. Since the data is already encoded into its representation, the footprint is very small when training the classifiers. Then, the generative filter for each user is trained separately and only on the single partition of personal data. Meanwhile, the adversarial and utility classifiers for each user are trained separately and only on the 8 parts of public data combined with the one part of personal data. The final result is 2 generative filters, one for each user, that correspond to their own choice of private and utility labels. After privatization through use of the user specific filters, we can evaluate the classification accuracy on the private and utility labels as measured by adversaries trained on the full privatized dataset, which is the combination of each users privatized data.

\subsection{More results of MNIST experiments}
\label{supp:mnist:more:res:w:mi}
\begin{figure}[!hpbt]
    \centering
    \includegraphics[width=0.55\textwidth]{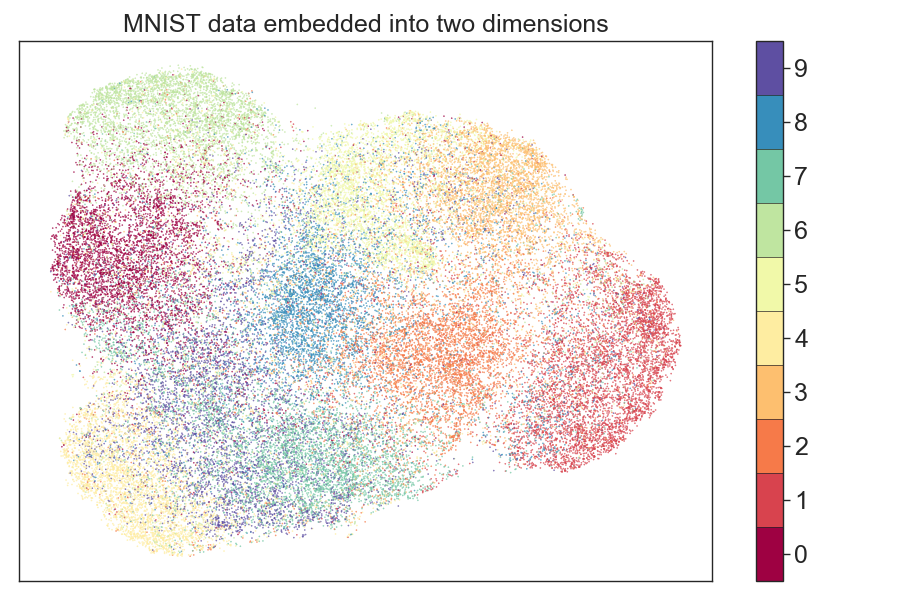}
    \caption{\textbf{MNIST case 2}: Visualization of the perturbed latent geometry. We discover that \textsf{0} is closer to \textsf{6}, compared with the original latent geometry in \Figref{fig:circle:latent:geo:raw}, a finding that indicates that it would be more difficult to distinguish which of those two digits is larger than or equal to five, even though both are even digits. }
    \label{fig:mnist:latent:geo:oddeven}
\end{figure}

In this section, we illustrate detailed results for the MNIST experiment when we set whether the digit is odd or even as the utility label and whether the digit is greater than or equal to 5 as the private label. We first show samples of raw images and privatized images in \Figref{fig:MNIST:samples:odd:even}. We show the classification accuracy and its sensitivity in \Figref{fig:MNIST:clf:acc:odd:even}. Furthermore, we display the geometry of the latent space in \Figref{fig:mnist:latent:geo:oddeven}. 

In addition to the classification accuracy, we evaluate the mutual information, to confirm that our generative filter indeed decreases the correlation between released data and private labels, as shown in \Figref{fig:mnist:mi:plot:case1:case2}.

\subsubsection{Utility of Odd or Even}
\begin{figure*}[!hpbt]
    \centerline{
    \begin{subfigure}[t]{0.51\columnwidth}
    \includegraphics[width=1\textwidth]{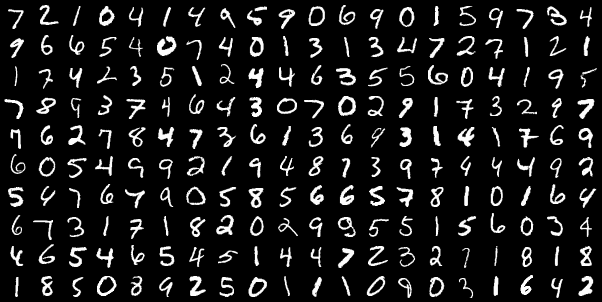}
    \caption{Sample of original digits}
    \end{subfigure}
    \begin{subfigure}[t]{0.51\columnwidth}
    \includegraphics[width=1\textwidth]{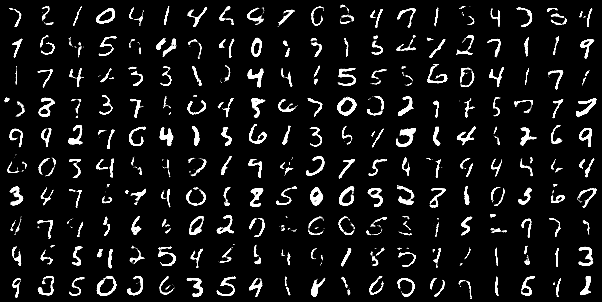}
    \caption{Same images large-valued digits privatized}
    \end{subfigure}
    }
    \caption{\textbf{MNIST case 2}: Visualization of digits pre- and post-noise injection and adversarial training. We discover that some large-valued digits ($\geq 5$) are randomly switched to low-valued ($<5$) digits (or vice versa) while some even digits remain even digits and odd digits remain as odd digits. }
    \label{fig:MNIST:samples:odd:even}
\end{figure*}

\begin{figure}[!hpbt]
    \centerline{
    \begin{subfigure}[t]{0.31\columnwidth}
    \includegraphics[width=1\textwidth]{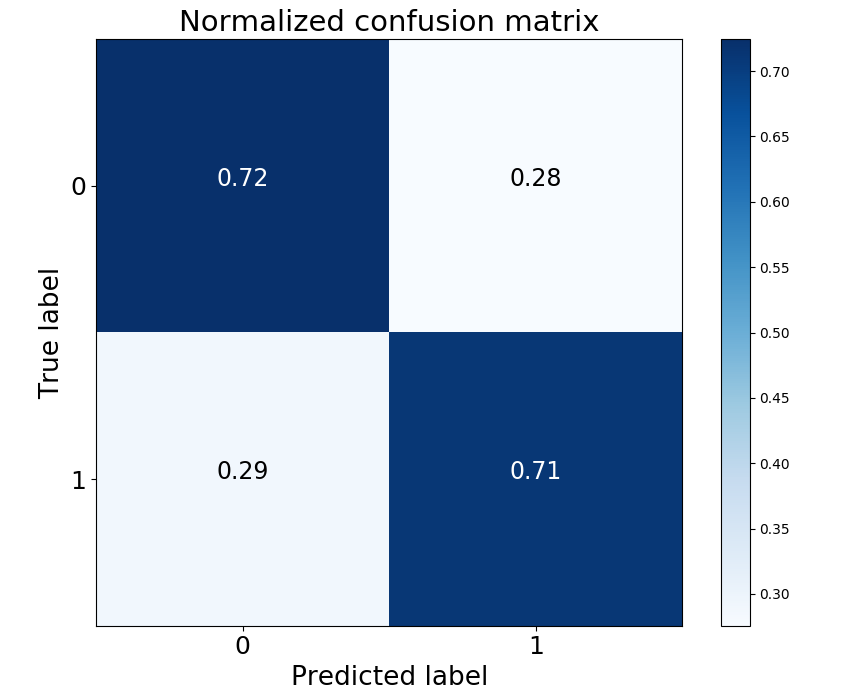}
    \caption{Confusion matrix of classifying the private label when $b=3$}
    \label{fig:mnist:confu:mat:priv:geq5:util:oddeven}
    \end{subfigure}
    ~
    \begin{subfigure}[t]{0.36\columnwidth}
    \includegraphics[width=1\textwidth]{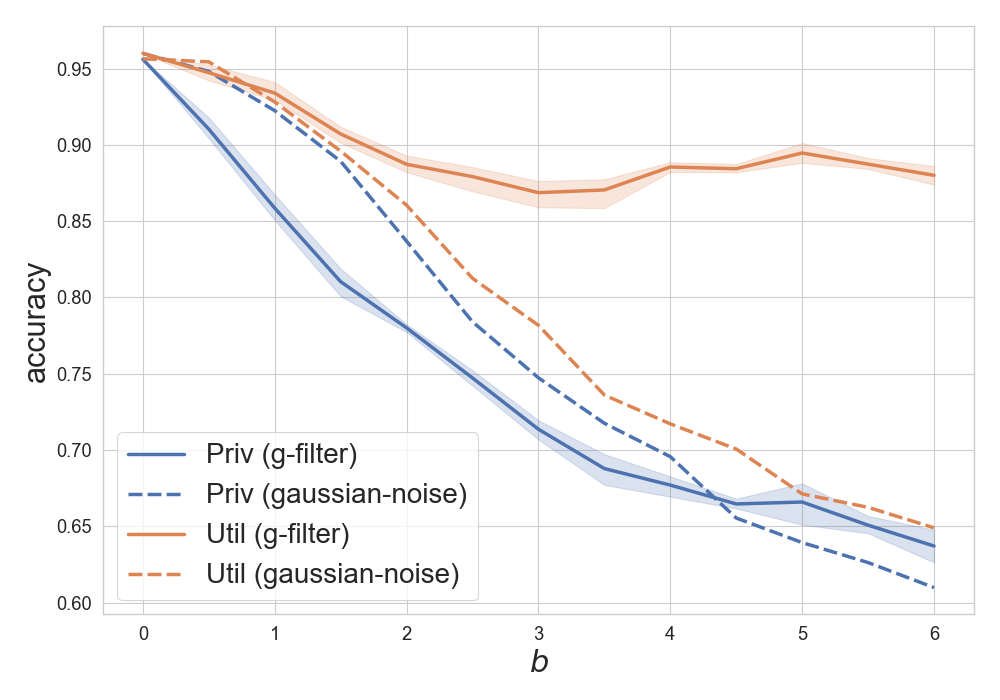}
    \caption{Sensitivity of classification accuracy versus the distortion $b$}
    \label{fig:mnist:acc:inc:b:priv:geq5:util:oddeven}
    \end{subfigure}
    ~
    \begin{subfigure}[t]{0.36\columnwidth}
    \includegraphics[width=1\textwidth]{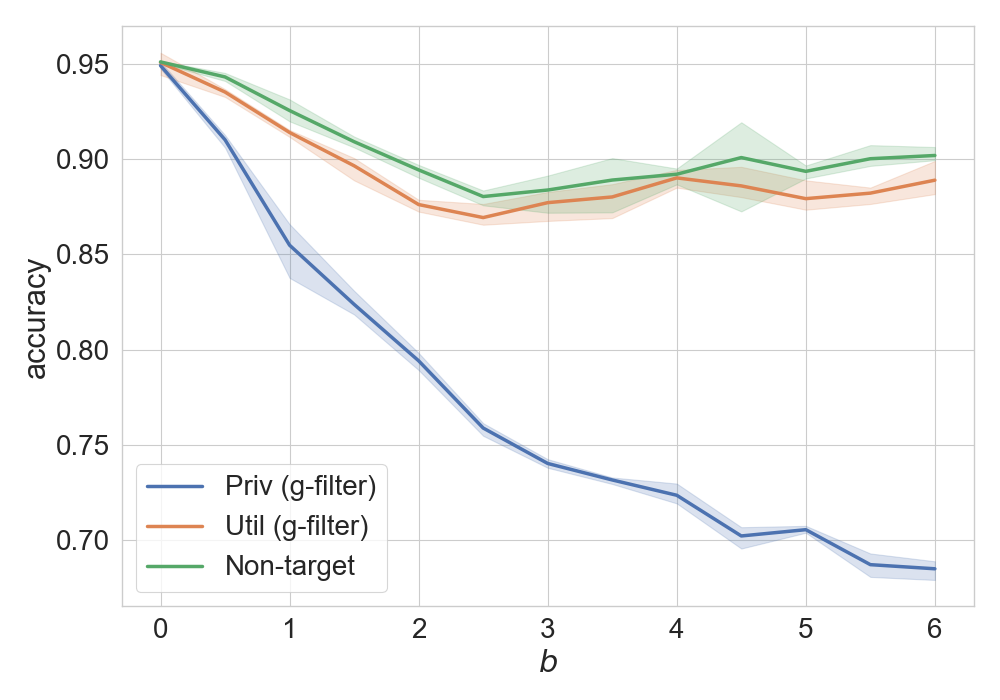}
    \caption{Sensitivity of classification accuracy versus the distortion $b$ with a non-target attribute}
    \label{fig:mnist:acc:inc:b:priv:geq5:util:non-target}
    \end{subfigure}
    }
    \caption{\textbf{MNIST case 2}: \Figref{fig:mnist:confu:mat:priv:geq5:util:oddeven} shows the false positive and false negative rates for classifying the private label when the distortion budget is 3 in KL-divergence. The \Figref{fig:mnist:acc:inc:b:priv:geq5:util:oddeven} shows that when we use the generative adversarial filter, the classification accuracy of private labels drops from 95\% to almost 65\% as the distortion increases, while the utility label can still maintain close to 90\% accuracy throughout. Meanwhile, the additive Gaussian noise performs worse because it yields higher accuracy on the private label and lower accuracy on the utility label, compared to the generative adversarial filter. \Figref{fig:mnist:acc:inc:b:priv:geq5:util:non-target} shows how non-target attributes not included in the filter's loss function (circle) can still be preserved due to the distortion budget restricting noise injection. The error bars show the standard error over a batch of 10 samples}
    \label{fig:MNIST:clf:acc:odd:even}
\end{figure}

We present some examples of digits when the utility is an odd or even number (\Figref{fig:MNIST:samples:odd:even}). The confusion matrix in \Figref{fig:mnist:confu:mat:priv:geq5:util:oddeven} shows that false positive rate and false negative rate are almost equivalent, indicating the perturbation resulting from the filter doesn't necessarily favor one type (pure positive or negative) of samples. \Figref{fig:mnist:acc:inc:b:priv:geq5:util:oddeven} shows that the generative filter, learned through minmax robust optimization, outperforms the Gaussian mechanism under the same distortion budget. The Gaussian mechanism reduces the accuracy of both private and utility labels, whereas the generative filter can maintain the accuracy of the utility while decreasing the accuracy of the private label, as the distortion budget goes up.

Furthermore, the distortion budget prevents the generative filter from distorting non-target attributes too severely. This budget allows the data to retain some information even if it is not specified in the filter's loss function. \Figref{fig:mnist:acc:inc:b:priv:geq5:util:non-target} shows the classification accuracy with the added non-target label of circle from MNIST case 1. 

\subsubsection{Empirical Mutual information}
\label{supp:mnist:more:mi}
\begin{figure}[!hbpt]
    \centerline{
    \begin{subfigure}[t]{0.45\columnwidth}
    \includegraphics[width=1\textwidth]{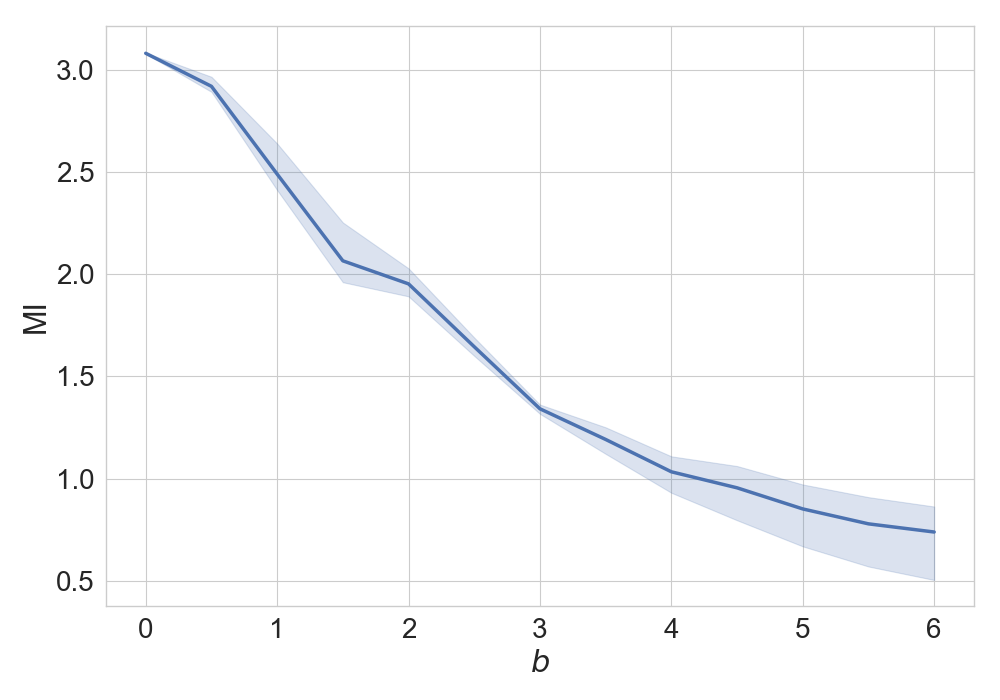}
    \caption{Digit identity as the private label}
    \end{subfigure}
    \begin{subfigure}[t]{0.45\columnwidth}
    \includegraphics[width=1\textwidth]{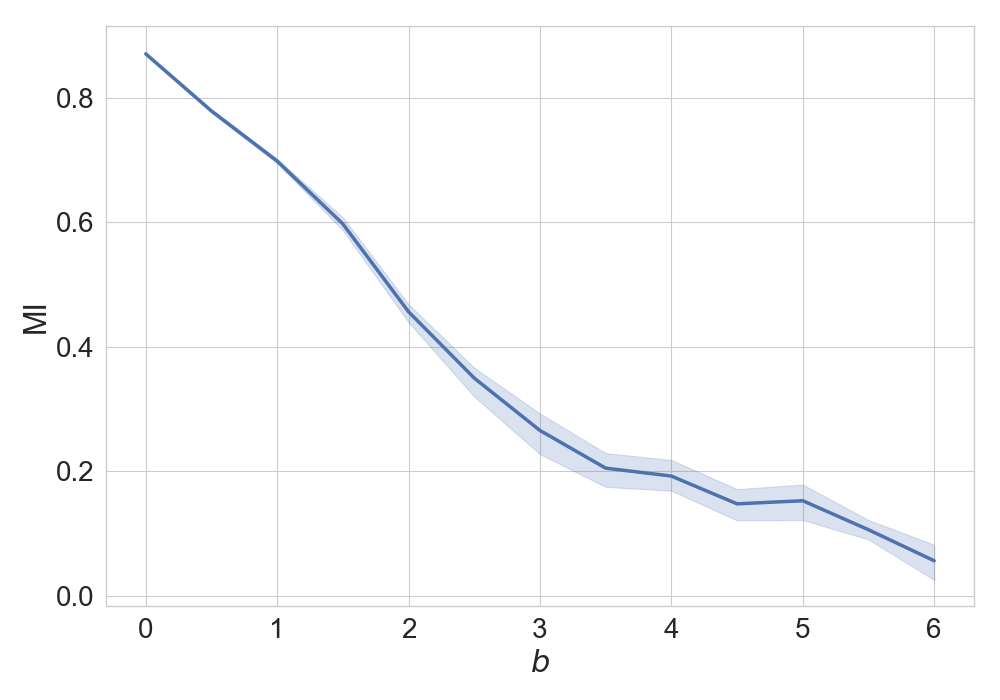}
    \caption{Large- or small-value as the private label}
    \end{subfigure}
    }
    \caption{Mutual information between the perturbed embedding and the private label decreases as the distortion budget grows, for both MNIST case 1 and case 2. }
    \label{fig:mnist:mi:plot:case1:case2}
\end{figure}

We use the empirical mutual information \citep{gao2015efficient} to verify if our new perturbed data is less correlated with the sensitive labels from an  information-theoretic perspective. The empirical mutual information is clearly decreased as shown in  \Figref{fig:mnist:mi:plot:case1:case2}, a finding that supports our generative adversarial filter can protect the private label given a certain distortion budget. 

%
%
\subsection{Additional information on CelebA experiments}
\subsubsection{Comments on comparison of classification accuracy}
We notice that our classification results on utility label (smiling) in the CelebA experiment perform worse than the state of the art classifiers presented in \cite{liu2015faceattributes} and \cite{torfason2016face}. However, the main purpose of our approach is not building the best image classifier, but getting a baseline of a comparable performance on original data without the noise perturbation. Instead of constructing a large feature vector (through convolution, pooling, and non-linear activation operations), we compress a facial image down to a 50-dimension vector as the embedding. We make the perturbation through a generative filter to yield a vector with the same dimensions. Finally, we construct a neural network with two fully-connected layers and an elu activation after the first layer to perform the classification task. We believe the deficit of the accuracy is because of the compact dimensionality of the representations and the simplified structure of the classifiers. We expect that a more powerful state of the art classifier trained on the released private images will still demonstrate the decreased accuracy on the private labels compared to the original non-private images while maintaining higher accuracy on the utility labels. This hypothesis is supported by the empirically measured decrease in mutual information demonstrated in \secref{supp:mnist:more:mi}.

\subsubsection{More examples of CelebA}
\label{appx:sec:celebA:examples:more}
\begin{figure*}[!hpbt]
    \centerline{
    \begin{subfigure}[t]{0.54\columnwidth}
    \includegraphics[width=1\textwidth]{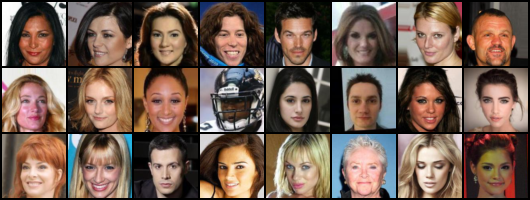}
    \end{subfigure}
    \begin{subfigure}[t]{0.54\columnwidth}
    \includegraphics[width=1\textwidth]{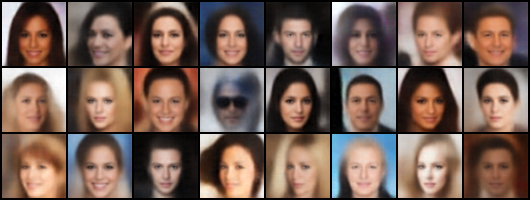}
    \end{subfigure}
    }
    \caption{Visualizing raw samples (on left) and encoded-decoded samples (on right) from a trained VAE with the Lipschitz smoothness.}
    \label{fig:celebA:raw:enc:dec:samples}
\end{figure*}

\begin{figure*}[!hbpt]
    \centerline{
    \begin{subfigure}[t]{0.54\columnwidth}
    \includegraphics[width=1\textwidth]{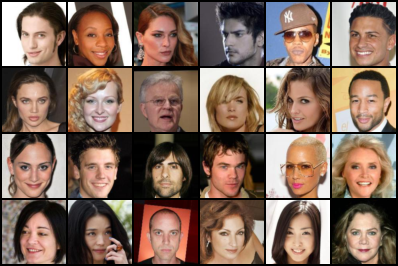}
    \caption{raw}
    \label{fig:celebA:samples:raw:attractive}
    \end{subfigure}
    \begin{subfigure}[t]{0.54\columnwidth}
    \includegraphics[width=1\textwidth]{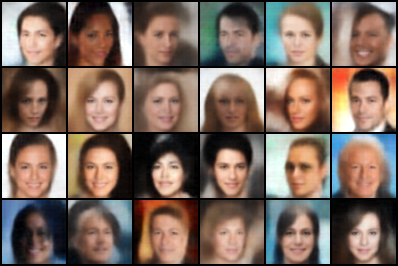}
    \caption{attractive privatized}
    \label{fig:celebA:samples:priv:attractive}
    \end{subfigure}
    }
    \caption{Sampled images. We find some non-attractive faces switch to attractive faces and some attractive looking images are changed into non-attractive, from left \Figref{fig:celebA:samples:raw:attractive} to right \Figref{fig:celebA:samples:priv:attractive}. }
    \label{fig:celebA:example:priv:attractive}
\end{figure*}

\begin{figure*}[!hbpt]
    \centerline{
    \begin{subfigure}[t]{0.54\columnwidth}
    \includegraphics[width=1\textwidth]{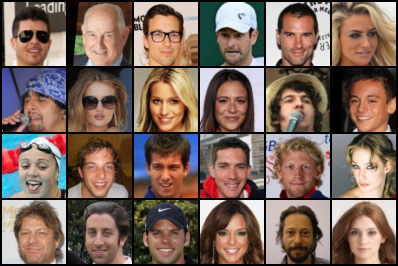}
    \caption{raw}
    \label{fig:celebA:samples:raw:eyegls}
    \end{subfigure}
    \begin{subfigure}[t]{0.54\columnwidth}
    \includegraphics[width=1\textwidth]{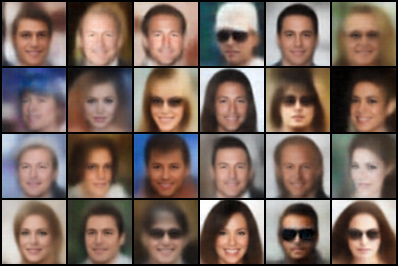}
    \caption{eyeglass privatized}
    \label{fig:celebA:samples:priv:eyegls}
    \end{subfigure}
    }
    \caption{Sampled images. We find some faces with eyeglasses are switched to non-eyeglasses faces and some non-eyeglasses faces are changed into eyeglasses-wearing faces, from left \figref{fig:celebA:samples:raw:eyegls} to right \figref{fig:celebA:samples:priv:eyegls}. }
    \label{fig:celebA:example:priv:eyeglasses}
\end{figure*}

\begin{figure*}[!hbpt]
    \centerline{
    \begin{subfigure}[t]{0.54\columnwidth}
    \includegraphics[width=1\textwidth]{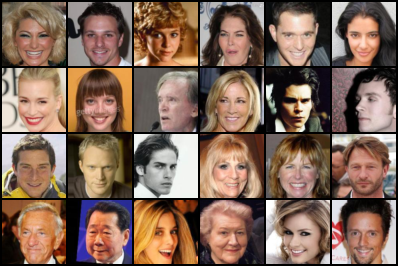}
    \caption{raw}
    \label{fig:celebA:samples:raw:wavehair}
    \end{subfigure}
    \begin{subfigure}[t]{0.54\columnwidth}
    \includegraphics[width=1\textwidth]{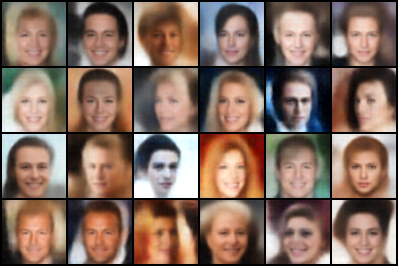}
    \caption{wavy hair privatized}
    \label{fig:celebA:samples:priv:wavehair}
    \end{subfigure}
    }
    \caption{Sampled images. We discover that some faces with wavy hair switch to images with non-wavy hair after our filter's perturbation (and vice versa), from left \figref{fig:celebA:samples:raw:wavehair} to right \figref{fig:celebA:samples:priv:wavehair}. }
    \label{fig:celebA:example:priv:wavyhair}
\end{figure*}

In this part, we illustrate more examples of CelebA faces yielded by our generative adversarial filter (Figures \ref{fig:celebA:example:priv:attractive}, \ref{fig:celebA:example:priv:eyeglasses}, and \ref{fig:celebA:example:priv:wavyhair}). We show realistic looking faces generated to privatize the following labels: attractive, eyeglasses, and wavy hair, while maintaining smiling as the utility label. The blurriness of the images is typical of state of the art VAE models because of the compactness of the latent representation \citep{higgins2017beta, dupont2018learning}. The blurriness is not caused by the privatization procedure but by the encoding and decoding steps as demonstrated in \Figref{fig:celebA:raw:enc:dec:samples}.

\subsection{Deferred equations}
In this section, we describe the $f$-divergence between two Gaussian distributions. The characterization of such a divergence is often used in previous derivations.  
\subsubsection{\texorpdfstring{\renyi} \text{ }divergence between Gaussian distributions} \label{sec:renyi:div:gauss}
$\alpha$-\emph{\renyi~divergence between two multivariate Gaussians} \big[\cite{gil2013renyi} and \cite{hero2001alpha}, Appendix Proposition 6\big]:
\begin{align}
    D_{\alpha}(p_i||p_j) = \frac{\alpha}{2} (\mu_i- \mu_j)^T \Sigma_{\alpha}^{-1}(\mu_i - \mu_j) - \frac{1}{2(\alpha-1)}\log\Big(\frac{|\Sigma_{\alpha}|}{|\Sigma_i|^{1-\alpha}|\Sigma_j|^{\alpha}}\Big), \label{app:deferred:eq:alph:div}
\end{align}
where $\Sigma_{\alpha} = \alpha\Sigma_j + (1-\alpha)\Sigma_i$ and $\alpha\Sigma_i^{-1}+(1-\alpha)\Sigma_j^{-1} > 0 $ is positive definite. \\
We give a specific example of two Gaussian distributions $p$ and $q$ that are $\mathcal{N}(\mu_1, \Sigma)$ and $\mathcal{N}(\mu_2, \Sigma)$ respectively. Letting $\E_{\mu_2}$ denote expectation over $x\sim \mathcal{N}(\mu_2, \Sigma)$, we then have 
\begin{align}
    & D_{\alpha}(p||q) =  \frac{1}{\alpha-1}\log\int\Big(\frac{p(x)}{q(x)}\Big)^{\alpha} q(x) dx \\ 
    & = \frac{1}{\alpha-1}\log \E_{\mu_2}\Big[\exp\Big(-\frac{\alpha}{2}(x- \mu_1)^T\Sigma^{-1}(x-\mu_1) + \frac{\alpha}{2}(x- \mu_2)^T\Sigma^{-1}(x-\mu_2)  \Big) \Big] \\
    & \explainEq{i}\frac{1}{\alpha-1} \log \E_{\mu_2}\Big[\exp\Big(-\frac{\alpha}{2}(\mu_1- \mu_2)^T\Sigma^{-1}(\mu_1-\mu_2) + {\alpha}(\mu_1- \mu_2)^T\Sigma^{-1}(x-\mu_2)  \Big) \Big] \\
    & \explainEq{ii} 
    \frac{1}{\alpha-1} \log\Big( \exp\Big[\frac{-\alpha}{2}(\mu_1- \mu_2)^T\Sigma^{-1}(\mu_1-\mu_2) + \frac{\alpha^2}{2}(\mu_1-\mu_2)^T\Sigma^{-1}(\mu_1 - \mu_2) \Big] \Big) \\
    & = \frac{1}{\alpha-1} \frac{(\alpha - 1)\alpha}{2}(\mu_1-\mu_2)^T\Sigma^{-1}(\mu_1 - \mu_2) \\
    & =\frac{\alpha}{2} (\mu_1-\mu_2)^T\Sigma^{-1}(\mu_1 - \mu_2), \label{app:deferred:renyi:two:gaussian:sameVar}
\end{align}
where equality (i) uses the relationship $-(x-a)^2+(x-b)^2 = -(a-b)^2 + 2(a-b)(x-b)$ and equality (ii) uses a linear transformation of Gaussian random variables $(\mu_1- \mu_2)^T\Sigma^{-1}(x-\mu_2) \sim \mathcal{N}\big(0, (\mu_1-\mu_2)^T\Sigma^{-1}(\mu_1 - \mu_2)\big)$.

\vfill


\end{document}